\documentclass[12pt,pdfstartview=FitB]{JHEP3}
\pdfoutput=1

\usepackage{amsmath,epsfig}
\usepackage{amssymb,amsfonts}
\usepackage{latexsym}
\usepackage{tocvsec2}
\usepackage{subeqnarray}
\usepackage{xcolor}

\usepackage{graphicx}
\usepackage{amsmath}
\usepackage{amssymb,amsfonts}
\usepackage{longtable}

\relax
\def\be{\begin{equation}}
\def\ee{\end{equation}}
\def\bea{\begin{eqnarray}}
\def\eea{\end{eqnarray}}

\newcommand\fverb{\setbox\pippobox=\hbox\bgroup\verb}
\newcommand\fverbdo{\egroup\medskip\noindent%
                        \fbox{\unhbox\pippobox}\ }
\newcommand\fverbit{\egroup\item[\fbox{\unhbox\pippobox}]}

\newcommand{\bear}{\begin{eqnarray}}

\newcommand{\eear}{\end{eqnarray}}

\newcommand{\bsea}{\begin{subeqnarray}}
\newcommand{\esea}{\end{subeqnarray}}
\newbox\pippobox

\def\6{\partial}

\def\a{\alpha}

\def\p{\partial}

\def\sq
\def\a{\alpha}

\def\hri#1#2{\href{http://arxiv.org/abs/#1}{[ArXiv:#1]#2}}
\def\hre#1#2{\href{http://arxiv.org/abs/#1/#2}{[ArXiv:#1/#2]}}

\newcommand{\ud}{\mathrm{d}}

\newcommand{\half}{\frac{1}{2}}

\allowdisplaybreaks[3]

\title{Universal scaling properties of extremal cohesive holographic phases}
\author{\large B. Gout\'eraux$^{a}$\\
~\\
~\\
$^a$\href{http://www.nordita.org}{Nordita}, KTH Royal Institute of Technology and Stockholm University\\
Roslagstullsbacken 23, SE-106 91 Stockholm, Sweden
\\\\
E-mail: \email{blaise@kth.se}
}

\preprint{NORDITA-2013-59}

\abstract{We show that strongly-coupled, translation-invariant holographic IR phases at finite density can be classified according to the scaling behaviour of the metric, the electric potential and the electric flux introducing four critical exponents, independently of the details of the setup. Solutions fall into two classes, depending on whether they break relativistic symmetry or not. The critical exponents determine key properties of these phases, like thermodynamic stability, the (ir)relevant deformations around them, the low-frequency scaling of the optical conductivity and the nature of the spectrum for electric perturbations. We also study the scaling behaviour of the electric flux through bulk minimal surfaces using the Hartnoll-Radicevic order parameter, and characterize the deviation from the Ryu-Takayanagi prescription in terms of the critical exponents.}

\keywords{AdS/CFT, AdS/CMT, holography, strong coupling, finite density, quantum criticality}

\begin{document}

\section{Motivations and results}
\label{intro}

Holographic applications to strongly-coupled Condensed Matter systems in the vicinity of a Quantum Critical Point have generated a flurry of activity over the recent years (for reviews, see \cite{Hartnoll:2011fn,Iqbal:2011ae}). A lot of attention has very naturally focussed on the Infra-Red (IR) dual geometries, which control many of the scaling properties of the low-temperature quantum critical phases. In particular, the IR scaling of thermodynamic quantities such as the entropy or the heat capacity, as well as transport coefficients such as the optical conductivity or the resistivity, have turned out to be largely independent from the microscopic details of the theory. This is an additional motivation to examine the IR phases more closely, as they provide some universal insight into the low-temperature dynamics, irrespective of the UV completion (which justifies adopting a UV conformal fixed point instead of say, a microscopic lattice).

It was proposed in \cite{cgkkm} to adapt the Wilsonian approach to effective field theories: effective holographic theories are then used to study and characterize the IR landscape of holographic strongly-coupled theories.  By selecting only the few most relevant operators in the IR, an effective action can be adopted which captures the essence of the IR dynamics
\be
\label{EMDaction}
S=\int \ud^{d+2}x\sqrt{-g}\left[\mathcal R-\frac{\partial\phi^2}2-\frac{Z(\phi)}4F^2-\frac{W(\phi)}2A^2+V(\phi)\right],\quad \left\{\begin{array}{c}Z(\phi)\underset{\phi\to\infty}{\to}Z_0 e^{\gamma\phi}\\ W(\phi)\underset{\phi\to\infty}{\to}W_0 e^{\varepsilon\phi} \\V(\phi)\underset{\phi\to\infty}{\to}V_0 e^{-\delta\phi}\end{array}\right.,
\ee
where we have retained the metric, a Maxwell field and a scalar (because of the mass term $W(\phi)$ this also encompasses the case where this scalar is the modulus of a charged complex scalar, like in holographic superfluids \cite{HoloSc}).

Moreover, we assume that the scalar coupling functions $Z$, $W$ and $V$ have exponential asymptotic branches when $\phi$ grows large, as is the case in all known gauged supergravities.
Translation-invariant phases preserving a U(1) symmetry in the IR ($W_0=0$) were examined first, \cite{cgkkm,gk}, followed by a study of symmetry-breaking ones ($W_0\neq0$),\footnote{Note however that a distinction should be made between (spontaneous) breaking of the U(1) symmetry in the UV and in the IR: if the symmetry-breaking terms are irrelevant in the IR, the U(1) symmetry can be restored there while still being (spontaneously) broken in the UV, see \cite{gk2012}.} \cite{gk2012}.

A classification of such phases is essential to the understanding of the low-temperature phase diagram at finite density. Indeed, the relevant question is: Shooting down from the UV, where does the (holographic) RG flow end, given a set of UV couplings (like a magnetic field, a scalar coupling, etc.)? To answer this fundamental question, it is necessary to have an overview of all possible IR phases in order to construct the allowed flows, carry out the thermodynamic analysis and determine the dominant phase.

In \cite{gk2012}, we have established such a classification for massive vector theories along three criteria: the scaling symmetries of the metric, the IR behaviour of the scalar and last, the presence of a constant electric flux or not. The most general scaling Ansatz for the metric (retaining translation invariance and homogeneity, and supported by a scalar diverging logarithmically in the IR) is
\be\label{HVLif}
	\ud s^2=r^{\frac2d\theta}\left(\frac{L^2\ud r^2+\ud x^2_{(d)}}{r^2}- \frac{\ud t^2}{r^{2z}}\right)
\ee
where $z$ and $\theta$ are the anisotropic dynamical and the hyperscaling violating exponent, respectively, \cite{cgkkm, gk, sachdev}. The former fixes the relative scaling between time and space
\be\label{LifScaling}
t\to\lambda^z t\,,\quad x_i\to \lambda x_i\,,\quad r\to\lambda r\,.
\ee
For nonzero hyperscaling violation exponent $\theta$, the metric \eqref{HVLif} is only left covariant under the scaling action \eqref{LifScaling}:
\be\label{CovMetric}
\ud s^2\to\lambda^{\frac2d\theta}\ud s^2\,.
\ee
This property can be linked to the low-temperature scaling of the thermal entropy, \cite{gk,sachdev}
\be\label{EntScaling}
S\sim T^{\frac{d_\theta}{z}},\qquad d_\theta=d-\theta
\ee
which displays some effective spatial dimensionality $d_\theta$: this is hyperscaling violation, measured by $\theta$. The results of \cite{gk} allow to interpret this effective dimensionality in a large region of the parameter space as the spatial dimensionality of a decompactified theory, via some Kaluza-Klein oxydation [as well as incidentally providing a resolution of possible IR curvature singularities of \eqref{HVLif}].\footnote{For other resolutions using magnetic fields, see \cite{MagIRres}.} In \cite{sachdev}, it was emphasized that for the specific value $\theta=d-1$, the system is effectively one-dimensional, as one would expect for a system with a Fermi surface. Moreover, the holographic entanglement entropy displays a logarithmic violation of its area law \cite{Ogawa:2011bz,sachdev,dong}, again along general expectations when a Fermi surface is present. As no IR bulk charges were in sight in the setup of \cite{Ogawa:2011bz,sachdev,dong}, this led to the conjecture that the 'fermionic' degrees of freedom responsible for the logarithmic violation must be 'hidden' behind the extremal horizon. However, \cite{semilocal} showed that current-current correlators exhibited no spectral weight at finite momentum and low energy, which seems to be in tension with the conjecture of \cite{Ogawa:2011bz,sachdev}.\footnote{Recent studies \cite{Edalati:2013tma} have also shown that systems with $\theta=d-1$ do not seem to display a zero sound mode.}

In any case, the holographic calculation of the entanglement entropy \cite{RyuTaka} only depends on (a constant time slice of) the IR geometry, and so cares nothing for the details of the theory, or for the behaviour of the time component of the metric or the gauge field. This means that while the original calculations \cite{Ogawa:2011bz,sachdev,dong} embedded the metric Ansatz \eqref{HVLif} with $\theta=d-1$ in the setup \eqref{EMDaction} without any U(1)-breaking mass term ($W_0=0$, with $W_0$ defined in \eqref{EMDaction}), the same logarithmic violation will be displayed in a theory with $W_0\neq0$, \cite{gk2012}. In this case, there are explicit charged degrees of freedom in the bulk (such as a superfluid condensate), with generically no electric flux emanating from the horizon. It is not known how current-current correlators would behave (along the lines of \cite{semilocal}), but one might conjecture that if there is to be a Fermi surface of hidden dofs, then it must be made up of dofs not charged under the broken U(1).

It seems therefore very important to sharpen the distinction between these two configurations. This can be understood by examining the contributions to the boundary charge density, \cite{Iqbal:2011in,Hartnoll:2011fn}: either from a charged IR horizon, or from charged bulk fields. This describes fractionalized or cohesive phases respectively, generalizing to finite density the usual interpretation in holography that event horizons are dual to deconfined phases \cite{Witten:1998zw}. Quantum phase transitions between fractionalized and cohesive hyperscaling violating phases were identified in \cite{Hartnoll:2011pp,Adam:2012mw,gk2012}, mediated by a scale-invariant ($\theta=0$) fixed point (dynamically) destabilized by a relevant (complex) deformation.

Clearly, the holographic entanglement entropy is oblivious to fractionalisation or cohesion. In \cite{Hartnoll:2012ux}, two observables were proposed:
\begin{itemize}
\item The first simply calculates the amount of flux $\Phi_\Gamma$ threading the bulk spatial hypersurface obtained by minimizing the area of the entangling region, that is the surface $\Gamma$ obtained via the Ryu-Takayanagi prescription. This bulk hypersurface is such that $\partial \Gamma=\Sigma$, where $\Sigma$ is the entangled region on the boundary of AdS. This observable is just thought of as additional data to the entanglement entropy, which would provide some measure of the density matrix of the degrees of freedom carrying the U(1) charge. It would be quite interesting to understand better what aspects of spacetime could be reconstructed given this data, along the lines of \cite{RecBulk}.
\item The second proposes to determine the bulk minimal hypersurface by minimizing a deformed entanglement entropy
\be\label{NewObs}
S_E^\lambda=\frac{A_\Gamma}{4G_N}+\lambda\Phi_\Gamma
\ee
where $A_\Gamma$ is the area of the minimal surface $\Gamma$ while $\Phi_\Gamma$ is the electric flux threading it, $\lambda$ a coupling between the two.
This new observable adds to the usual entanglement entropy prescription a dipole coupling to the electric field. This could happen on D-branes with an internal dipole moment, polarized under some external electric field. They are conjectured to be dual to a class of surface operators, though no precise identification of the field theory observables has been provided yet.
\end{itemize}
Assuming the region $\Sigma$ to be composed of two parallel $(d-1)$-dimensional hypersurfaces separated by a distance $L$ along the remaining spatial direction, fractionalised phases were found to display a volume scaling law
\be
\Phi_\Gamma=\rho\, vol(\Sigma)L
\ee
where $\rho$ is the boundary charge density. For partially fractionalised phases, only the coefficient changes and should be replaced by $\rho-\rho_c$, where $\rho_c$ accounts for charge density sitting outside the deep IR region.

\cite{Hartnoll:2012ux} carried out the analysis for cohesive phases with a constant charge density $\sigma_{eff}(r)=\sigma_0$ in the IR:
\be\label{GenMaxEqIR}
\nabla_\mu\left(Z(\phi)F^{\mu\nu}\right)=j^\nu_{eff}\,,\qquad j_{eff}^\nu=\sigma_0 u^\nu\,,
\ee
where we have allowed for a non-mininal coupling $Z(\phi)$ between a scalar and the electric field, while $u^\nu$ is a timelike unit vector.\footnote{One can also consider confined phases, where the IR geometry terminates at some cut-off radius, but we shall not do so here.} The scaling of the new observable \eqref{NewObs} at large $L$ was found to be dominated by the usual area term if $\theta\leq0$, just like the entanglement entropy, while for $\theta>0$, the flux term $\Phi_\Gamma$ would be more important. In particular, this means that when the term due to the electric flux dominates, the minimal surface found differs from the Ryu-Takayanagi prescription, providing a nonlocal order parameter sensitive to cohesion.

\paragraph{Summary and results\\}

In this work, we shall provide evidence that there exists a universal parameterization of translation-invariant, cohesive phases. Extending the results of \cite{gk2012}, we will assume that electric flux conservation is broken explicitly in the IR and that Gauss's law takes the form
\be\label{GenMaxEqIR2}
\nabla_\mu\left(Z(\phi)F^{\mu\nu}\right)=j^\nu_{eff}\left(\phi, A^\lambda\right)\,,\qquad j^\nu_{eff}=\sigma_{eff}(r) u^\nu\,.
\ee
The field strength can couple non-minimally to a neutral scalar via the coupling $Z(\phi)$, and the electric potential typically couples to a current $j^\nu_{eff}\left(\phi, A^\lambda\right)$, which can depend on the scalar but also on the vector field. This allows to account for several cases of interest, where the charge density $\sigma_{eff}$ might originates from a U(1)-breaking term (which is a substitute for a charged condensate in the bulk, \cite{gk2012}), from a parity-violating term (whereby the magnetic field generates electric charge and can lead to stripe instabilities, \cite{Donos:2011bh,Donos:2012yu,Donos:2013wia,Withers:2013loa}) or from a fermionic fluid in the bulk, \cite{ElectronStar,Hartnoll:2011pp}. Letting $\sigma_{eff}$ depend on the radial coordinate, this leads us to introduce two new scaling exponents. The first, $\xi$, parametrises the IR scaling of the electric flux
\be\label{GenScalingE}
E(r)=\int Z(\phi)\star F\underset{IR}{\sim}r^{\xi}\,,
\ee
which is of course is related to the presence or absence of fractionalised degrees of freedom in the IR. For that reason, we call it the \emph{cohesion exponent}. The fractionalised limit corresponds to $\xi=0$.

The second, $\zeta$, parametrises the violation of Lifshitz scaling in the electric potential
\be\label{GenScalingA}
A_t\underset{IR}{\sim}r^{\zeta-\xi-z}\ud t
\ee
together with the cohesion exponent $\xi$ when the phase is cohesive. Anticipating a little, we call it the \emph{conduction} exponent.

 The various scaling exponents $z$, $\theta$, $\xi$ and $\zeta$ are determined by the parameters of the Lagrangian. This parameterization holds irrespective of the behaviour of the scalar field in the IR, which typically can either go to a constant or run logarithmically. It allows to lift the degeneracy between various solutions for which the metric is Lifshitz invariant ($\theta=0$) which have appeared in previous literature: solutions with a constant scalar and a massive vector field (or $p$-form) studied in \cite{taylor,Li} are cohesive but preserve Lifshitz scale invariance with $\zeta=\xi=-d$; on the other hand, fractionalised Lifshitz solutions with a running scalar \cite{taylor, KT} have $\xi=0$ and $\zeta=-d_\theta$, while their cohesive version still has $\zeta=-d_\theta$ but $\xi\neq0$, \cite{gk2012,kasa,nbi}, and generically break Lifshitz invariance.

Fractionalised solutions are obtained when the source term on the right-hand side of \eqref{GenMaxEqIR2} vanishes and coincide with the solutions of \cite{cgkkm}. Partially fractionalised solutions are obtained when this source term is irrelevant in the IR, and so Gauss's law is restored at leading order.

In section \ref{section1}, we exhibit large families of (often novel) solutions to the setups mentioned above. Remarkably, they fall into two classes, which are distinguished by whether the current dual to the electric potential is relevant or not in the IR:
\begin{itemize}
\item If it is relevant, it breaks Poincar\'e symmetry and thus time scales anisotropically compared to space with $z\neq1$. The value of the conduction exponent is fixed to $\zeta=-d_\theta$. The IR background is then charged.
\item If it is irrelevant, Poincar\'e symmetry is restored with $z=1$ and time and space scale identically. The value of the conduction exponent is then independent from the values taken by the other scaling exponents. In this case, one needs to carefully distinguish whether it is consistent or not to switch off the electric potential in Maxwell's equation. If yes the IR background is effectively neutral and an exact solution to the field equations (this is the case for the massive vector theories, where the electric potential enters on both sides of Maxwell's equation, or in the presence of a Chern-Simons coupling); if not, the electric potential parameterizes a background power series (this is the case for the electron stars). In our notation of \eqref{GenMaxEqIR2}, the two cases are distinguished by whether $j^\nu_{eff}\left(\phi, A^\lambda\right)$ vanishes as $A^\lambda$ is turned off or not.
\end{itemize}

In order to get a better theoretical handle on the cohesion and conduction exponents $\xi$ and $\zeta$, we proceed to study the scaling of various observables in the field theory. The IR dimension of operators can be determined by working out the static, purely radial deformations around the solution. We find that they come by pairs, which one expects to sum to the universal value $d_\theta+z$ on general dimensional grounds by inserting some operator $\int\ud t\,\ud^{d_\theta} x\, g_{\mathcal O}\,\mathcal O$ in the IR field theory in $d_\theta$ spatial dimensions. This expectation is not realised when the electric potential sources a background power series as well as the deformation: in this case, the value of the sum is shifted to $1-\zeta$ in $d=2$,\footnote{for which we have solutions, but we expect a similar expression to generalise in arbitrary $d$.} and is controlled by the conduction exponent $\zeta$. It is also not borne out when the electric potential is irrelevant and coupled to a magnetic component via a Chern-Simons coupling: in this case, the sum is $2+\zeta-\theta$. Moreover, there is always a marginal mode $\beta_0=0$ conjugate to a relevant mode $\beta_u=d_\theta+z$, which respectively are rescalings of time and turning on a small nonzero temperature. The other modes are non-universal (they depend on the details of the theory) and the dual dimensions will typically display all scaling exponents $z$, $\theta$, $\xi$ and $\zeta$.

In section \ref{section:AC}, we will show that the scaling exponents govern the low-frequency scaling of (the real part of) the optical conductivity at zero temperature. We find that
\be
\begin{split}
z\neq1\,,\,\zeta=-d_\theta\,: &\qquad Re(\sigma)\sim\omega^{2-\frac2z+\frac{d_\theta}z}\,,\\
z=1\,,\,\zeta\neq-d_\theta\,: &\qquad Re(\sigma)\sim\omega^{-\zeta}\,,
\end{split}
\ee
which leads us to conjecture that it should scale generically as
\be
 z\neq1\,,\,\zeta\neq-d_\theta\,:\qquad Re(\sigma)\sim\omega^{2-\frac2z-\frac{\zeta}z}
\ee
for generic $z$ and $\zeta$, although such phases remain to be discovered. This shows that the conduction exponent $\zeta$ determines the scaling of the optical conductivity. The scalings above are valid when $T\ll\omega\ll\mu$ where $T$ and $\mu$ are the temperature and chemical potential of the dual theory. The $z=1$ scaling is identical to the scaling of the conductivity for probe charge carriers in a neutral IR background, even though our phases are always charged (when $z=1$ the current is irrelevant but nonzero).

Imposing the Null Energy Condition and thermodynamic stability, we show that the optical conductivity always vanishes at zero frequency (as it should in order for linear response theory to be valid), while the spectrum of electric perturbations is gapless. In the thermodynamically unstable region, see figure \ref{fig:spectrum}, there is a small parameter space where this is still true, but more generally linear response theory breaks down, and the spectrum is gapped if the Schr\"odinger potential diverges in the UV. This is always the case in $d>2$, and in $d=2$ if the UV dimension of the scalar obeys $1/2<\Delta<1$. The expectation is then that there will be a phase transition at low temperatures before that IR solution is reached. 

Finally, motivated by the analysis of \cite{Hartnoll:2012ux}, we will show in section \ref{section3} that the value of the cohesion exponent $\xi$ controls the scaling of the proposed order parameter \eqref{NewObs} for a strip of width $L\gg1$:
\be
\begin{split}
L\gg1\,,\quad \xi\leq-d_\theta\,,&\qquad S_E^\lambda\sim A_\Gamma\sim L^{1-d_\theta}\\
L\gg1\,,\quad \xi>-d_\theta\,,&\qquad S_E^\lambda\sim \Phi_\Gamma\sim  L^{\frac{2-d_\theta+\xi}{2-d_\theta-\xi}}
\end{split}
\ee
For $\xi\leq-d_\theta$, the area term always dominates the new observable \eqref{NewObs}, which means that the scaling is identical to the Ryu-Takayanagi prescription in the deep IR. However, some care is needed to determine the dynamics of the hypersurface, which may differ sensibly from the RT prescription. Otherwise, the flux term dominates, and the minimal surface is different from the Ryu-Takayanagi prescription. Interestingly, the limiting value is identical to the value taken by the conduction exponent in $z\neq1$ phases: then, both terms contribute equally, and the electric potential \eqref{GenScalingA} is Lifshitz-invariant (though the metric need not be).

We conclude in section \ref{section:conclusions}.

\section{A universal parameterization of cohesive phases\label{section1}}

In this section, we will show that the parameterization \eqref{GenScalingE} and \eqref{GenScalingA} together with \eqref{HVLif} holds for a variety of models, which display cohesive phases in the IR: models with a massive vector, a fermion fluid in the bulk or with a Chern-Simons term. This way, we realise three different ways to neutralise the extremal horizon.

We will assume that there is a scalar coupling to the other fields. In the IR, it can either settle in an extremum of its effective potential or run along a logarithmic branch. In that case, the couplings are approximated by exponentials for convenience, following intuition from supergravity potentials. In what follows, we will only consider explicitly the runaway case, but it is straightforward to see that our results also hold for the first case. In order to derive the hyperscaling solutions, one simply needs to replace in the solutions the amplitudes of the various exponential couplings $V_0$, $Z_{0,1,2}$, $W_0$ or $\vartheta_0$ by the values taken by these couplings at the extremum of the effective scalar potential where the scalar settles.\footnote{This trick does not work for the deformations, which have to be worked out from scratch.}

We will scan through three different models representing three different options to generate electric flux in the bulk in the presence of a neutral extremal horizon, i.e. with vanishing IR electric flux. We will find that we can organise the discussion along two broad classes: non-relativistic solutions with $z\neq1$ but a fixed conduction exponent $\zeta=\theta-d$; and relativistic solutions with $z=1$ and independent conduction exponent $\zeta$. Whether one or the other is reached depends on the nature of the operator dual to the vector field in the IR (relevant or irrelevant). Finally, we examine static, purely radial deformations around our solutions and show that the conjugate pairs generically sum to $d_\theta+z$, though their details depend on the model and also involve $\zeta$, $\xi$. This does not hold when the current is irrelevant and the electric flux is sourced by a fermion fluid or a magnetic field: in those cases, a pair of deformations sums anomalously.

\subsection{The models\label{section:models}}

\subsubsection{Massive vectors\label{section:MassEMDsol}}

In \cite{gk2012}, the following class of theories were studied in four-dimensional bulk spacetimes $d=2$
\be
\label{EMDaction2}
S=\int \ud^{d+2}x\sqrt{-g}\left[\mathcal R-\frac{\partial\phi^2}2-\frac{Z_0 e^{\gamma\phi}}4F^2-\frac{W_0 e^{\varepsilon\phi}}2A^2+V_0 e^{-\delta\phi}\right],
\ee
which might be thought of as an effective theory for the asymptotics of a holographic superfluid (when the U(1) is broken spontaneously).

The field equations read:
 \be
\begin{split}
&R_{\mu\nu} + \frac{Z}{2} \, F_{\mu\rho} F^{\rho}_{\;\;\nu} -  \half \p_\mu \phi \, \p_\nu \phi - \frac{W}{2} A_\mu A_\nu\\
&\qquad\qquad +\frac{g_{\mu\nu}}{2} \left[\half(\p \phi)^2 - V -R   +  \frac{W}{2} \, A^2 +  \frac{Z}{4}  F^2  \right] = 0 , \;\;\; \\
\label{scalar0}
&\Box \phi   = \frac{1}{4} Z'(\phi) \, F^2 + \half W'(\phi) A^2-V'(\phi)  \, , \\
&\frac{1}{\sqrt{-g}} \, \p_\mu \left(\sqrt{-g} \, Z(\phi) \, F^{\mu\nu}\right) = W(\phi) A^\nu  \, .
\end{split}
\ee
For power-like solutions, it was found that if $\epsilon\neq\gamma-\delta$, the mass term on the right of \eqref{GenMaxEqIR2} (here $j_\mu\left(\phi, A^\lambda\right)=W_0 e^{\epsilon\phi}A_\mu$) is irrelevant in the IR and vanishes at leading order in $r$, giving rise to partially fractionalised phases. On the other hand, if $\epsilon=\gamma-\delta$, then the mass term contributes at leading order in $r$ in Maxwell's equation, resulting in cohesive phases where the electric flux vanish in the IR. We will recall these solutions in section \ref{section:Solutions} and generalise them to arbitrary $d$. Note that as mentioned in the introduction, the source vanishes when the vector is switched off.

\subsubsection{Electron stars}

Turning to fermions, \cite{ElectronStar} proposed that a density of bulk fermions could be modeled using a Thomas-Fermi approximation. This allows to go beyond the probe limit, and introduces an extra term in the Lagrangian, equal to the pressure of the fermionic fluid, $p(\mu_{loc})$. $\mu_{loc}$ is the local chemical potential felt by the fermions at radial coordinate $r$, with $\mu_{loc}^2=g^{\mu\nu}A_\mu A_\nu$.\footnote{Defined as the time component of the gauge field in the orthonormal frame, \cite{ElectronStar}.} Solutions with a macroscopic density of fermions carrying all of the electric charge were constructed numerically, and named electron stars. If the mass of the fermions is too large compared to the chemical potential, the Fermi sea is not populated and a fractionalised phase can be found, with all the electric flux emanating from behind the extremal horizon. Here, we will relax the requirement for a constant density of fermions, and find new possible IR asymptotics for the electron stars.

We will consider the effective four-dimensional Lagrangian
\be
\frac{\mathcal L_{ES}}{\sqrt{-g}}=\frac1{2k^2}\left(R-\frac12\partial\phi^2+V_0 e^{-\delta\phi} \right)-\frac1{4e^2} e^{\gamma\phi}F^2-e^{\epsilon\phi}p(\mu_{loc})
\ee
which gives rise to the following effective field equations for the ideal fluid of charged fermions:
\be\label{FieldEqES}
\begin{split}
&G_{\mu\nu}=\frac12\partial_\mu\phi\partial_\nu\phi-\frac14\partial\phi^2g_{\mu\nu}+\frac{k^2}{e^2}e^{\gamma\phi} \left(F_{\mu\rho}F_\nu{}^\rho-\frac{1}{4}F^2g_{\mu\nu}\right)+\frac12V_0 e^{-\delta\phi}g_{\mu\nu}+k^2e^{\epsilon\phi} T_{\mu\nu}\\
&\nabla_\mu\left(e^{\gamma\phi}F^{\nu\mu}\right)=e^2 e^{\epsilon\phi} j^\nu\\
&\Box\phi=\delta V_0 e^{-\delta\phi}+\gamma\frac{k^2}{e^2}e^{\gamma\phi}F^2-2k^2\epsilon e^{\epsilon\phi} p
\end{split}
\ee
where $e$ is the gauge coupling and $k=1/8\pi G_N$. Note that we have allowed for a direct coupling between the scalar and the fermion fluid.
The perfect fluid tensor reads
\be
T_{\mu\nu}=(\rho+p)u_\mu u_\nu +p g_{\mu\nu}
\ee
with energy density $\rho$ and pressure $p$, while the charge current\footnote{which differs from $j^\mu_{eff}\left(\phi, A^\lambda\right)$ introduced in \eqref{GenMaxEqIR2} by a factor of $e^{\epsilon\phi}$. Observe that it is not consistent to switch off the vector without switching off the electron star as well.} is
\be
j_\mu=\sigma u_\mu
\ee
where $\sigma$ is the charge density of the fermions and $u_\mu$ is a unit vector. 
We will define their dimensionless version
\be
p=\frac{\hat p}{k^2}\,,\qquad \rho=\frac{\hat\rho}{k^2}\,,\qquad \sigma=\frac{\hat\sigma}{ek^2}\,.
\ee
 They can be expressed as integrals of the density of states
\be\label{densitiesES}
\hat\rho=\hat\beta\int^{k\mu_{loc}/e}_{\hat m}\epsilon^2\sqrt{\epsilon^2-\hat m^2}\ud\epsilon\,,\quad \hat\sigma=\hat\beta\int^{k\mu_{loc}/e}_{\hat m}\epsilon\sqrt{\epsilon^2-\hat m^2}\ud\epsilon\,,\quad -\hat p=\hat\rho-\frac{k}e\mu_{loc}\hat\sigma
\ee
where
\be
\hat m^2 = \frac{k^2}{e^2}m^2\,,\qquad \hat\beta=\frac{e^4}{k^2\pi^2}\,.
\ee
They also verify the first law
\be
\hat p'=\left(\frac{k}e\mu_{loc}\right)'\hat\sigma\,.
\ee

In \cite{ElectronStar}, the focus was on electron stars with a constant charge density in the IR, where a Lifshitz scaling could be seen to emerge. Here, we would like to investigate the possibility of violating hyperscaling. Since these geometries need a logarithmically running scalar to support them, we expect the local chemical potential to run as well, which means it becomes very large compared to the fermion mass, $\mu_{loc}\gg m$.\footnote{We will check in retrospect in which parameter space this approximation is valid.} This allows to simplify the integrals in \eqref{densitiesES} to obtain:
\be\label{densitiesESHyp}
\begin{split}
\hat\sigma=\frac{\hat\beta}{3}\left(\frac{k\mu_{loc}}{e}\right)^3-\frac{\hat m^2}{2}\left(\frac{k\mu_{loc}}{e}\right)+O(\hat m^4),\\
 \hat\rho=\frac{\hat\beta}{4}\left(\frac{k\mu_{loc}}{e}\right)^4-\frac{\hat m^2}{4}\left(\frac{k\mu_{loc}}{e}\right)^2+O(\hat m^4)  ,\\
 \hat p =\frac{\hat\beta}{12}\left(\frac{k\mu_{loc}}{e}\right)^4 -\frac{\hat m^2}{4}\left(\frac{k\mu_{loc}}{e}\right)^2+O(\hat m^4) 
\end{split}
\ee
As advertised, since the first term in the expansion is mass-independent, the fermions are effectively massless in the IR, where $\mu_{loc}\to+\infty$.

Such solutions where the local chemical potential is scaling in the IR have not been constructed in previous literature (at least for generic $z$), and it would be interesting to do so.  If $\epsilon=2\gamma-\delta$, we will exhibit a cohesive solution \eqref{EstarCoh} with arbitrary dynamical exponent $z\neq1$ and fixed conduction exponent $\zeta=\theta-2$, while if $\epsilon\neq2\gamma-\delta$, a cohesive solution with fixed dynamical exponent $z=1$ and arbitrary conduction exponent $\zeta\neq\theta-2$ is obtained \eqref{EstarCohz=1}.

\subsubsection{Phases with Chern-Simons couplings}

Let us now turn to models still preserving the U(1) symmetry in the IR, but including a parity violating term, which can generate phases with charge density waves and spatial modulation, \cite{Donos:2011bh,Donos:2012yu,Donos:2013wia,Withers:2013loa}:
\be
S=\int\ud^4x\sqrt{-g}\left(R-\frac12\partial\phi^2-\frac{Z_1(\phi)}{4}F_1^2-\frac{Z_2(\phi)}{4}F_2^2+V(\phi)+\frac14 \vartheta(\phi) F_{1}\tilde F_2\right)
\ee
with
\be
Z_1(\phi)\underset{IR}{\sim}Z_1e^{\gamma_1\phi},\quad Z_2(\phi)\underset{IR}{\sim}Z_2e^{\gamma_2\phi},\quad V(\phi)\underset{IR}{\sim}V_0e^{-\delta\phi},\quad \vartheta(\phi)\underset{IR}{\sim}\vartheta_0e^{\lambda\phi}.
\ee
and
\be
\tilde F^{\kappa\lambda}=\frac12\epsilon^{\kappa\lambda\mu\nu}F_{\mu\nu}\,,\qquad \epsilon^{\kappa\lambda\mu\nu}=\frac{1}{\sqrt{-g}}\varepsilon^{\kappa\lambda\mu\nu}\,.
\ee
When the two gauge fields are the same, this model coincides with the model of \cite{Donos:2011bh,Donos:2012yu,Donos:2013wia,Withers:2013loa}. The field equations read
\be
\begin{split}
&2R_{\mu\nu}=\partial_\mu\phi\partial_\nu\phi+\sum_{i=1,2}Z_i(\phi)\left(F_{i\mu}{}^\rho F_{i\nu\rho}-\frac{g_{\mu\nu}}4F_i{}^2\right)-V(\phi)g_{\mu\nu}\\
&\nabla_\mu\left(Z_{1}(\phi)F_{1}^{\mu\nu}-\frac12\vartheta(\phi)\tilde F_{2}^{\mu\nu}\right)=0,\qquad \nabla_\mu\left(Z_{2}(\phi)F_{2}^{\mu\nu}-\frac12\vartheta(\phi)\tilde F_{1}^{\mu\nu}\right)=0\\
&0=\Box\phi+V'(\phi)-\frac14Z_1'(\phi)F_1{}^2-\frac14Z_2'(\phi)F_2{}^2+\frac{\vartheta'(\phi)}{4}F_{1\kappa\lambda}\tilde F_2^{\kappa\lambda}
\end{split}
\ee
and the $F_1\tilde F_2$ term does not appear in Einstein's equations because of the $\sqrt{-g}$ factor in the definition of the Levi-Civita tensor. As for the electron stars, the effective mass term in Maxwell's equation for $A_1$ only depends on $A_2$ (and vice and versa), which means that it is not consistent to switch off only one of the two vectors.

We will look for hyperscaling violating solutions where we only turn on a time component for the $A_1$ field and a spatial one for $A_2$
\be
 A_1=Q_0r^{\zeta-\xi-z}\ud t\,,\qquad A_2=\frac{Q_2}2\left(y\ud x-x\ud y\right).
\ee
which means that we are placing the dual field theory at finite density in a constant external magnetic field.

\subsection{Non-relativistic solutions \label{section:Solutions}}

\subsubsection{Massive vectors}

 Cohesive phases all the way to the IR can be found if $\epsilon=\gamma-\delta$,\footnote{As mentioned before, partially fractionalised phases are obtained if this relation does not hold.} and as advertised, two cases must be distinguished depending on whether the current is relevant or not. If it is, solutions with broken Poincar\'e symmetry $z\neq1$ can be found:
\be\label{MassiveEMDsol}
\begin{split}
&\ud s^2=r^{\frac2d\theta}\left(\frac{L^2\ud r^2+\ud R_{(d)}^2}{r^2}-\frac{\ud t^2}{r^{2z}}\right),\qquad  A=Q_0r^{\zeta-\xi-z}\ud t\,,\qquad  e^\phi=r^{\pm\kappa}\,,\\
& L^2V_0=(d-1+z-\theta)(d+z-\theta)+(z-1)\xi\,,\qquad  Q_0^2= \frac{2 (z-1)}{Z_0(z-\zeta+\xi )}\,,\\
&\kappa=\sqrt{2 (1-z) (\zeta-\xi) +\frac2d\theta(\theta-d) }\,,\qquad \delta\kappa =\pm \frac{2\theta }{d}\\
&\frac{W_0}{Z_0}= -\frac{(z-\zeta+\xi )\xi}{L^2}\,,\qquad \gamma\kappa = \pm2\left(\frac1d\theta- \zeta +\xi\right)\,,\qquad \epsilon=\gamma-\delta\,.
\end{split}
\ee
In this solution, $\zeta=\theta-d$ is fixed. The electric flux scales like \eqref{GenScalingE}, and setting $\xi=0$ recovers fractionalised phases. Two different Lifshitz solutions exist: one with $\theta=0$, which is a cousin of those studied in \cite{KT}, while another exists for $\xi=\zeta=\theta-d$, where it is now the electric potential rather than the metric which is Lifshitz invariant. Setting further $\theta=0$ recovers a pure Lifshitz solution with a constant scalar, \cite{Li,taylor}.

There is a locally critical limit, taking $(z,\theta,\xi)\to+\infty$ while keeping the ratios $\theta/z$, $\xi/z$ finite, which provides solutions conformal to AdS$_2\times \mathbf R^{d}$:
\be\label{MassiveEMDsolSL}
\begin{split}
&\ud s^2=r^{\frac2d\theta}\left(-\frac{\ud t^2}{r^{2}}+\frac{L^2\ud r^2}{r^2}+\ud R_{(d)}^2\right),\qquad L^2V_0=(1-\theta)^2+\xi\\
&A=Q_0r^{\theta-\xi-1}\ud t\,,\qquad Q_0^2= \frac{2}{Z_0(1-\theta+\xi )}\,,\qquad  e^\phi=r^{\pm\kappa}\,,\qquad \kappa=\sqrt{2\xi +\frac2d\theta(\theta-d) }\\
&\frac{W_0}{Z_0}= -\frac{(1-\theta+\xi ) \xi}{L^2}\,,\qquad \delta\kappa =\pm \frac{2\theta }{d}\,,\qquad \gamma\kappa = 2\pm\left(\frac\theta{d}-\theta+\xi \right),\qquad \epsilon=\gamma-\delta\,.
\end{split}
\ee
Their fractionalised version $\xi=0$ has been studied previously, \cite{semilocal} and displays special Kaluza-Klein properties \cite{gk}. It would be interesting to find out whether their cohesive generalisations display different properties.

The limit $\theta=0$ can be taken in \eqref{MassiveEMDsolSL} provided $\xi\geq0$, but at the price that $W_0<0$. This seems in tension with the interpretation of $W_0$ as a charge squared, but would open up the possibility of having a spacetime which is AdS$_2\times\mathbf R^d$ with flux $\xi=0$ or without flux $\xi<0$. This would effectively decorrelate the fact that an infinite throat opens up in the IR from the behaviour of the electric flux. As far as we are aware, cohesive AdS$_2\times\mathbf R^d$ ground states have not be considered previously in the literature.

\subsubsection{Electron stars}

 Plugging the leading order expressions for the fluid quantities \eqref{densitiesESHyp} in the field equations \eqref{FieldEqES}, we look for power-like solutions. Assuming that all powers of $r$ contribute equally implies $\epsilon=2\gamma-\delta$, we find:\footnote{There is also a solution which has $\theta=2+z$, but its radial, static deformations are logarithmic, so we discard it.}
\be\label{EstarCoh}
\begin{split}
&\ud s^2=r^{\theta}\left(-\frac{\ud t^2}{r^{2z}}+\frac{L^2\ud r^2+\ud R_{(2)}^2}{r^2}\right),\qquad  A=\frac{k}{e}Q_0r^{\theta-z-2-\xi}\ud t\,,\qquad e^\phi=r^\kappa\\
&Q_0=\sqrt{\frac{z-1}{2+z-\theta +\xi }},\qquad\qquad \hat\beta =\frac{3 \xi  (2+z-\theta +\xi )^2}{(1-z)  L^2}\\
&L^2=\frac{ \left(4+2 z^2-6 \theta +2 \theta ^2-\xi +z (6-4 \theta +\xi )\right)}{2 V_0}
\end{split}
\ee
with
\be
\gamma =\frac{4-\theta +2 \xi}\kappa\,,\quad \delta = \frac{\theta}\kappa,\quad \epsilon=\gamma-2\delta,\quad \kappa=\sqrt{2(1-z)(\theta -2-\xi )+\theta (\theta -2)}\,.
\ee
The scaling exponents $z$, $\theta$ and $\xi$ are determined in terms of the parameters of the theory, namely $\delta$, $\gamma$ and $\hat\beta$. Our solution is exact in the zero fermion mass case, but will receive corrections from the mass otherwise. $\xi$ is defined through the scaling of the electric flux as in \eqref{GenScalingE}. The conduction exponent $\zeta=\theta-2$ on the other hand is fixed, and can be read off from \eqref{GenScalingA}. 

 The fractionalisation limit is $\xi=0$, for which the fluid vanishes from the deep IR ($\hat\beta=0$). There is also a locally critical limit $(z,\theta)\to+\infty$, $\theta=-\eta z$ as previously, with $\eta$ now parameterizing the conformal deviation from AdS$_2\times\mathbf R^2$.

 In the limit in which we are working, the fermion mass does not enter at leading order in the IR.
As advertised, the fermion density and pressure also scale with $r$, since
\be
\mu_{loc}=\frac{e}kQ_0 r^{\frac{1}{2} (\theta-4 -2 \xi)}
\ee
Note that assuming that the local chemical potential blows up in the IR places some constraints on the parameter space: $(\theta-4 -2 \xi)(\theta-2)<0$. This is enough to keep the corrections at non-zero fermion mass $\hat m$ under control.
When $\theta-4 -2 \xi=0$, or equivalently $\kappa\gamma=0$, the local chemical potential and the charge density become constant in the IR, so that the solution is only valid for strictly massless fermions $\hat m=0$. In this limit, either the gauge coupling $Z(\phi)$ becomes constant or $\kappa=0$ (constant scalar). In the latter case, we recover a Lifshitz IR, which is the massless limit of the solution of \cite{ElectronStar}.

\subsubsection{Phases with Chern-Simons couplings}

Similarly to section \ref{section:MassEMDsol}, an analysis of possible IR phases in terms of relevant operators can be carried out. Looking at power solutions of the field equations, it is readily seen that unless the relation $2\lambda=\gamma_1+\gamma_2\neq0$ holds in the IR, the parity violating term is irrelevant at leading order, and the phase is partially fractionalised.

Assuming this relation to hold, cohesive hyperscaling violating solutions can be found:
\be\label{ParityViolEMDsol}
\begin{split}
&\ud s^2=r^\theta\left(-\frac{\ud t^2}{r^{2z}}+\frac{L^2\ud r^2+\ud x^2+\ud y^2}{r^2}\right),\quad L^2=\frac{(2+z-\theta)(1+z-\theta)}{V_0}\\
& Q_0=-\frac{L\vartheta_0Q_2}{2Z_1(\theta-2-\xi-z)},\qquad Q_2{}^2=\frac{8Z_1(z-1)(2+z-\theta)}{L^2(\vartheta_0^2+4Z_1Z_2)}\\
&  e^\phi=r^{\pm\kappa}\,,\qquad \kappa=\sqrt{(\theta-2)(\theta+2-2z)}\\
& \delta=\pm\frac\theta\kappa\,,\qquad \gamma_1=\pm\frac{4-\theta+2\xi}{\kappa}\,,\qquad \gamma_2=\pm\frac{\theta-4}{\kappa}\,,\qquad 2\lambda=\gamma_1+\gamma_2\,.
\end{split}
\ee
This solution with $z\neq1$ again verifies $\zeta=\theta-2$.
The electric flux scales like \eqref{GenScalingE} and vanishes in the IR unless $\xi=0$, in which case it is constant and the phase is fractionalised. This implies $\lambda=0$ and $\gamma_1=-\gamma_2$, which turns off the effects of the $F_1\wedge F_2$ term: there is nolonger a source on the right-hand side of Maxwell's equation, the magnetic flux nolonger carries electric charge, we generically have a dyonic solution. Solving for $\theta_0$, we obtain the following relation between $Q_0$ and $Q_2$:
\be
Q_2^2=\frac{2 (-1+z) V_0}{(1+z-\theta ) Z_1 Z_2}-\frac{(z-\zeta )^2 Q_0^2 V_0 Z_1}{(1+z-\theta ) (2+z-\theta ) Z_2}
\ee
We can now truncate it to a purely electric or magnetic one by setting either $Q_2=0$ or $Q_0=0$. This coincides then with single-gauge field model of \cite{Donos:2012yu}: since they have $\lambda=0$, their model can only admit purely magnetic solutions (or purely electric, fractionalised ones). On the other hand, for other values of $\xi$, all the bounday electric flux is sourced by the magnetic flux in the bulk.

Another truncation of this theory to a cohesive phase with a single gauge field is when both gauge fields can be regrouped in a single one: $\gamma_1=\gamma_2\equiv\gamma$ and $Z_1=Z_2$, which implies in turn $\zeta=2$ and $\lambda=\gamma$.

Finally, let us note the presence of a truncation of \eqref{ParityViolEMDsol} to a cohesive Lifshitz solution when $\theta=0$, as well as a locally critical limit $z,\,\theta,\,\xi\to+\infty$, which yields a solution conformal to AdS$_2\times\mathbf R^2$:
\be\label{ParityViolEMDsolSL}
\begin{split}
&\ud s^2=r^{\theta}\left(-\frac{\ud t^2}{r^{2}}+\frac{L^2\ud r^2}{r^2}+\ud x^2+\ud y^2\right),\qquad L^2=\frac{(1-\theta )^2}{V_0}\,,\qquad e^\phi=r^{\pm\kappa},\\
&z=1\,,\qquad Q_0^2=\frac{2 (1-\theta )\vartheta_0^2}{(\theta-\xi-1 )^2 Z_1^2 \left(\vartheta_0^2+4 Z_1 Z_2\right)}\,,\qquad  Q_2^2= \frac{8 (1-\theta )}{ L^2\left(\vartheta_0^2+4 Z_1 Z_2\right)}\\
&\kappa^2=\theta(\theta-2)\,,\quad \kappa\delta =\pm \theta\,,\quad \kappa\gamma_1 = \mp\left(\theta+2 \xi \right),\quad \kappa\gamma_2=\pm \theta\,,\quad \kappa\lambda=\pm\frac\xi2\,.
\end{split}
\ee
Note that since $\zeta=\theta$, the limit $\theta\to0$ recovers a dyonic AdS$_2\times\mathbf R^2$ solution with a vanishing (constant) scalar. However we may have $\zeta=\xi=\theta$ (the electric potential is then Lifshitz-invariant) while still allowing for a running scalar and hyperscaling violation $\theta\neq0$.

\subsection{Relativistic solutions}

Relativistic solutions with $z=1$ can be obtained by assuming that terms not coming from the metric or the scalar do not enter at leading order in $r$ in the field equations. This either means that the electric potential does not backreact on the metric and scalar field and only sources one of the deformations, like for the massive vectors or the dyonic solutions, or that it parameterizes a power series, like for the electron stars.

The scaling exponents $\zeta$ and $\xi$ are determined by matching to the parameterization \eqref{GenScalingA} and \eqref{GenScalingE}, and solving Maxwell equation independently from the others at leading order in the deformation or in the power series.
$\zeta$ will now appear as an independent parameter, contrarily to the $z\neq1$ solutions above, but this comes with having to set $z=1$.
The fractionalised limit is still $\xi=0$,\footnote{Such an example of relativistic, fractionalised phase with hyperscaling violation appears in section 8 of \cite{gk}.} but unsurprisingly it is distinct from its $z\neq1$ cousin (which would have $\zeta=\theta-d$).

\subsubsection{Massive vectors}

In this setup, the electric potential can consistently be switched off in Maxwell's equation, since the mass term on the right-hand side then vanishes. As a consequence, there is no background power series as in the next two subsections (though there is one consistent with a partially fractionalised phase, see \cite{gk2012}). The electric potential does not enter in the background, only  through the deformations:
\be\label{EMDsolCoh1}
L^2=\frac{(d+1-\theta)(d-\theta)}{V_0}\,,\quad \kappa=\sqrt{\frac2d\theta(\theta-d)}\,,\quad \kappa\delta=\pm\frac{2\theta}{d}\,,\quad Q_0=0\,.
\ee

\subsubsection{Electron stars}

It is clear from the expressions of $Q_0$ and $\hat\beta$ in \eqref{EstarCoh} that this solution is not valid for $z=1$. Analysing the equations of motion in that case, there are two possibilities: either an exact solution with $\theta=3$, or a power series if one assumes that terms coming from the gauge field and the fermion fluid contribute at subleading order in the IR in the scalar and Einstein's equations.\footnote{Treating the electric potential as a deformation around a neutral gravitational background would only yield a partially fractionalised phase, since the right-hand side of Maxwell equation would only matter at higher order in $\mu_{loc}$ given by \eqref{densitiesESHyp}.} The exact solution has the unfortunate property that the static, radial perturbations around it which should generate temperature are now logarithmic and so that solution seems pathological. The details of the power series solution read however
\be\label{EstarCohz=1}
\begin{split}
&\ud s^2=-D(r)\ud t^2+B(r)\ud r^2+C(r)\ud R^2_{(2)}\,,\quad A_t=A_0(r)\ud t\,,\quad \phi=\phi(r)\\
&D(r)=r^{\theta-2}\left[1+\sum_{n=1}d_nr^{n\alpha}\right],\quad B(r)=L^2r^{\theta-2}\left[1+\sum_{n=1}b_nr^{n\alpha}\right]\\
&C(r)=r^{\theta-2},\quad A_0(r)=Q_0 r^{\zeta-\xi-1}\left[1+\sum_{n=1}a_nr^{n\alpha}\right],\quad\phi(r)=\kappa\log r+\sum_{n=1}\Phi_nr^{n\alpha}\\
&\kappa\delta=\theta\,,\qquad \kappa\gamma=2-\zeta +2 \xi\,,\qquad \kappa\epsilon=2-3 \zeta +4 \xi\,,\qquad \kappa=\sqrt{\theta(\theta-2)}\\
&Q_0^2 =-\frac{3\xi  (1-\zeta +\xi )}{\hat\beta L^2}\,,\qquad L^2=\frac{(\theta-2)(\theta-3)}{V_0}\,,\qquad \alpha=\zeta+2-\theta\,.
\end{split}
\ee
where the $a_n$, $b_n$, $c_n$, $d_n$ and $\Phi_n$ are the coefficients of the power series and are uniquely determined by the field equations, typically proportional to $\hat\beta^{-1/2}$ emphasizing that the non-zero density of the fluid is the source of the background power series. $\alpha$ should always be such that terms going like $r^{n\alpha}$ with $n>0$ some integer decay faster towards the IR than $r^0$. The reason why there is a power series is that it is not consistent to switch off entirely the electric potential, which is sourced by the fluid charge density in the IR.\footnote{An alternative would be to assume a partially fractionalised phase as in \cite{Hartnoll:2011pp}, such that the fluid stops at some radius in the IR and hovers above the extremal horizon.}

The conduction exponent $\zeta$ must be different from $\theta-2$ in order for the power series to be well-defined, which is equivalent to $\epsilon\neq2\gamma-\delta$. $\epsilon$ is otherwise unfixed. The local chemical potential can be evaluated to be
\be
\mu_{loc}=Q_0\frac{k}e r^{\zeta -\frac{\theta }{2}-\xi}
\ee
which imposes some constraint on the parameter space in order to ensure it blows up in the IR: $(\theta-2)(2\zeta -\theta-2\xi)<0$. This is enough to ensure that the corrections to next order in the fermion mass $\hat m$ are indeed subleading.

A similar solution was constructed by \cite{Hartnoll:2011pp}, in section 3.4 for a model with $\gamma=\delta=1/\sqrt3$ and $\epsilon=0$. Note that $\gamma=\delta$ implies that the local chemical potential is constant and nolonger scales with $r$. More generically, this will happen whenever $\epsilon=\gamma-\delta$. Our results will then only coincide with those of \cite{Hartnoll:2011pp} in the strictly massless limit $\hat m=0$.

\subsubsection{Phases with Chern-Simons couplings}

The leading solution is precisely identical to \eqref{EMDsolCoh1}, with $Q_2=0$ as well. Both the electric and magnetic fields are irrelevant, and will be sourced through the deformations. Note that it would be possible to treat only the magnetic field as a deformation, and then the phase would be partially fractionalised with an electrically charged extremal background and $z\neq1$.

\subsection{Deformations}

We will only concern ourselves with purely radial perturbations at zero frequency and zero momentum. They are important to determine if the solution can be reached with a stable or RG flow or not. Deformations give the dimension of the dual corresponding operator in the IR, and come by pairs, which on general grounds from dimensional analysis (taking into account the effects of hyperscaling violation) should sum to $d+z-\theta$. Interestingly, we shall see that this expectation is not always borne out. When the charge current is irrelevant ($z=1$ and $\zeta\neq\theta-d$) and coupled either to a charged fluid of fermions or to a magnetic field via a Chern-Simons term, the deformations sourced by the electric potential sum anomalously to $1-\zeta$ (electron stars) or $2+\zeta-\theta$ (magnetic field), where the deviation is parameterized by $\zeta$.

In all the solutions exposed in the previous sections, one always finds a marginal mode $\beta_0=0$ corresponding to a rescaling of time and a constant shift of the scalar, as well as its conjugate $\beta_{u}=d+z-\theta$, which is relevant and puts the solution at nonzero temperature. These are universal in the sense that they do not depend on the details of the setup and are completely fixed by our metric Ansatz. 

Below, we detail the other modes $\beta_\pm$, which are both non-trivial and setup-dependent. We shall see however that they depend on the values of the cohesion and conduction exponents, ($\xi$, $\zeta$), and consequently, so do the dimensions of the IR operators they are dual to.

Finally, we give the allowed parameter space, where we request that: the solution is well-defined (no complex parameters, the power series when it exists is subleading); the electric flux vanishes in the IR; the solution is thermodynamically stable; the deformations are irrelevant; and for the electron stars, the local chemical potential diverges in the IR.

\subsubsection{Massive vectors}

We will parameterize the deformations in this section with the Ansatz\footnote{We fix the radial gauge by choosing $\Delta C(r)/C(r)=0$.}
\be
\frac{\Delta D(r)}{D(r)}=D_1 r^{\beta_\pm},\quad \frac{\Delta B(r)}{B(r)}=B_1 r^{\beta_\pm},\quad \frac{\Delta \phi(r)}{\phi(r)}=\Phi_1 r^{\beta_\pm},\quad \frac{\Delta A_t(r)}{A_t(r)}=A_1 r^{\beta_\pm^a}.
\ee

For simplicity, we give results for $d=2$, but the values for generic $d$ can be obtained by rescaling
\be
\left(z,\theta,\zeta,\xi\right)\to\frac{2}d\left(z,\theta,\zeta,\xi\right),\qquad \beta\to\frac{d}2\beta
\ee

\paragraph{Non-relativistic solution \protect\eqref{MassiveEMDsol}\\}

In this case, the last pair of modes around \eqref{MassiveEMDsol} reads:
\be
\begin{split}
&\beta^a_\pm=\beta_\pm=\frac12(d+z-\theta)\pm\frac12\sqrt{\frac{X_m}{2(1-z)(\theta-2 -\xi )+\theta (\theta -2)}}\\
&X_m=(2-\theta ) \left(16 z^3-34-32 z^2 (\theta-1 )+47 \theta -16 \theta ^2+2 z \left(8 \theta ^2-7-8 \theta \right)\right)\\
&\quad+2 (z-1) \xi  \left(81+72 z+8 z^2-96 \theta -36 z \theta +28 \theta ^2-64 \xi -24 z \xi +32 \theta  \xi -16 \xi ^2\right)
\end{split}
\ee
As shown in \cite{gk2012} for $d=2$, $\beta_+$ is never irrelevant, while the condition that $\beta_-$ is irrelevant is not always verified (it can also be relevant and/or complex).

\paragraph{Relativistic solution \protect\eqref{EMDsolCoh1}\\}
 
 In this case, the electric potential does not backreact on the metric and scalar at leading order in the IR, finite density occurs by turning it on as a deformation. If we parameterize the leading IR behaviour of the electric potential according to \eqref{GenScalingA}, this defines
\be
\beta_-^a=\zeta-\xi-1\,,\quad\beta_+^a=-\xi\,,\quad \epsilon=\gamma-\delta\,,\quad  \kappa\gamma=2\xi-\zeta+d-\theta+\frac2d\theta\,,\quad \frac{W_0L^2}{Z_0}=(\zeta-\xi-1)\xi\,.
\ee
Note that this is sourced only by $A_1$ with all other coefficients set to zero, since they can only mix at quadratic order. Ensuring that they do allows to fix the value of $\beta_\pm$ relative to $\beta^a_\pm$:
\be
\beta_\pm=\frac12(3-\theta)\pm\frac12(\zeta-1)=\beta_\pm^a+\frac12(4+2\xi-\theta-\zeta)\,,
\ee
This will be valid for the region of the parameter space where $(2-\theta)(\zeta+2-\theta)>0$, otherwise there are no irrelevant deformations.

The full parameter space is
\be
\theta <0\,,\quad\,\xi <0\,,\quad\,\zeta <\theta-2\,.
\ee
The condition on $\zeta$ is necessary to ensure that $\beta_-$ is irrelevant.

Note that none of these pairs sum anomalously, that is we always find that $\beta_++\beta_-=2+z-\theta$.

\subsubsection{Electron stars}
We will parameterize the deformations in this section with the Ansatz\footnote{We fix the radial gauge by choosing $\Delta C(r)/C(r)=0$.}
\be\label{DefnsES}
\begin{split}
\frac{\Delta D(r)}{D(r)}=D_1 r^{\beta_\pm}\sum_{n=1}d'_nr^{n\alpha}\,,\qquad \frac{\Delta B(r)}{B(r)}=B_1 r^{\beta_\pm}\sum_{n=1}b'_nr^{n\alpha}\,,\\
\frac{\Delta \phi(r)}{\phi(r)}=\Phi_1 r^{\beta_\pm}\sum_{n=1}\Phi'_nr^{n\alpha}\,,\qquad \frac{\Delta A_t(r)}{A_t(r)}=A_1 r^{\beta_\pm}\left[1+\sum_{n=1}a'_nr^{n\alpha}\right].
\end{split}
\ee
The power series is turned off for the non-relativistic solutions \eqref{EstarCoh} but is necessary for the relativistic solution \eqref{EstarCohz=1} where there is also a power series in the background.

\paragraph{Non-relativistic solution \protect\eqref{EstarCoh} \\}

We find in this case:
\be
\begin{split}
\beta_\pm=&\frac12\left(2+z-\theta\pm\sqrt{\frac{X}{\theta ^2-2 (2+\xi )+z (4-2 \theta +2 \xi )}}\right),\\
X=&(2+z-\theta ) (2-\theta ) \left(2 z+18 z^2+16 \theta -19 z \theta +\theta ^2-20\right)\\
&+2 \left(8 z-76+59 z^2+9 z^3+100 \theta -42 z \theta -34 z^2 \theta -41 \theta ^2+21 z \theta ^2+4 \theta ^3\right) \xi \\
&+8 \left(10 z-15+5 z^2+8 \theta -6 z \theta -\theta ^2\right) \xi ^2+32 (z-1) \xi ^3
\end{split}
\ee
The two modes $\beta_\pm$ sum to $2+z-\theta$, as expected.

Given other constraints on the parameter space, $\beta_-$ is always irrelevant (never relevant or complex).
The parameter space reads
\be
2<\theta <4\,,\quad\,\frac{1}{2} (\theta-4 )<\xi <0\,,\quad\,z<0\,.
\ee

\paragraph{Relativistic solution \protect\eqref{EstarCohz=1} \\}

There, we get
\be
\begin{split}
 &\beta_\pm=\frac{1 }{2}(1-\zeta)\pm\frac12\sqrt{1+\zeta ^2-8 \xi -8 \xi ^2+\zeta  (-2+8 \xi )}
\end{split}
\ee
As advertised, $\beta_++\beta_-=1-\zeta\neq2+z-\theta$. We interpret this deviation from the fact that the electric potential, while irrelevant, still participates in the background geometry (it cannot be turned off with the background still solving the field equations), and this shifts the dimensions of the dual operators. Note that the power series does not backreact exactly in the same way on the electric potential and on the metric or scalar in \eqref{DefnsES}, which makes the identification of the dual dimensions subtle. However, it is straightforward to see that the limit $\zeta\to\theta-2$ would recover the usual value.

The parameter space in that case is
\be
\theta <0\;\&\;\left(\xi \leq\theta -3\; \&\;\zeta <1+\xi \quad\|\quad\theta-3 <\xi <0\;\&\;\zeta <\theta-2 \right)\,.
\ee
Note that $\beta_-$ could be relevant and even complex given other constraints, and so requiring $\beta_-<0$ reduces the parameter space available.

\subsubsection{Phases with Chern-Simons couplings}

Magnetic fields are topological, and in the boundary dual theory are associated with a source but not with a vev, which means that the IR geometry should only allow for one free integration constant, which varies the value of the magnetic field in which the background is placed. In the IR, this is done by shifting the scalar by a constant, which corresponds to a marginal(ly relevant) mode. This is different from the case of \cite{MetalIns}, where radial deformations can be turned on without spoiling homogeneity, thanks to the Bianchi VII symmetry of the Ansatz. Here, no such radial deformations are allowed by homogeneity.

Turning to the charge, the pair of modes sourced by the electric potential display a shift when the current is irrelevant.

\paragraph{Non-relativistic solution \protect\eqref{ParityViolEMDsol} \\}
We parameterize the remaining deformations as
\be
\begin{split}
\frac{\Delta D(r)}{D(r)}=D_1 r^{\beta_\pm},\quad \frac{\Delta B(r)}{B(r)}=B_1 r^{\beta_\pm},\quad \frac{\Delta \phi(r)}{\phi(r)}=\Phi_1 r^{\beta_\pm},\quad \frac{\Delta A_t(r)}{A_t(r)}=A_1 r^{\beta_\pm}.
\end{split}
\ee

There is only one zero mode $\beta_0=0$ and $\beta_u=2+z-\theta$, coming from the metric. The other pair of modes is still sourced by the metric ($A_1=0$):
\be
\beta_\pm^{met}=\frac12\left(2+z-\theta\mp\frac{\sqrt{(\theta+2-2z)(\theta-2-z)(\theta^2-19z\theta+16\theta-20+18z^2+2z)}}{\theta+2-2z}\right)
\ee
$\beta_-^{met}$ is always irrelevant given the constraints from the background solution.
The electric potential sources a pair of modes (which would reduce to $\beta_0$ and $\beta_u$ if we turned off the source in Maxwell's equation, that is $\xi=0$):
\be
\beta_\pm^{el}=\frac12(2+z-\theta)\mp\frac12(2+z-\theta-2\xi)
\ee
Here, $\beta_+^{el}$ is always relevant, while $\beta_-^{el}$ is not always irrelevant in the allowed parameter space.

The parameter space is a bit too lengthy to write, but allows for a large range of values of $\theta$, $z$ and $\xi$.

\paragraph{Relativistic solution \protect\eqref{ParityViolEMDsolz=1} \\}

There are two pairs of modes summing to $3-\theta$: two zero modes $\beta_0=0$ (one of which is time rescaling, while the other is a shift of the scalar which can be used to vary the magnetic field $Q_2$) and associated nonzero temperature modes $\beta_u=2+z-\theta$ (compared to the non-relativitic solution, $\beta^{met}_\pm$ have collapsed to $\beta_0$ and $\beta_u$),. Then one more pair $\beta^{el}_\pm$ is associated to the electric charge. One is a zero mode which just shifts the electric potential by a constant (the $U(1)$ is not broken), while the other is written

\be\label{ParityViolEMDsolz=1}
\begin{split}
&\frac{\Delta \Phi}{\Phi}=\sum_{i,j=1}\Phi_{i,j} r^{i\beta_-^{el}+j(\beta_-^{el}+\psi+\theta-\zeta)} \qquad \Phi=B(r),D(r),\phi(r)\\
&A_1(r)=Q_0 r^{\zeta-\xi-1}\left[1+\sum_{i,j=1}A_{i,j} r^{i\beta_-^{el}+j(\beta_-^{el}-\psi+\theta-\zeta)}\right],\\
&A_2=(1+\xi-\zeta)\frac{Q_0}{L\theta_0}\left(y\ud x-x\ud y\right)\\
&  \kappa\lambda = \xi\,,\quad \kappa\gamma_1=2\xi-\zeta+2\,,\quad \kappa\gamma_2=\theta-2-\psi\,,\quad \beta_-^{el}=\zeta+2-\theta\,
\end{split}
\ee
where $\theta$, $\zeta$, $\xi$ and $\psi$ are all free parameters. $Q_0$ is the integration constant associated to the mode, and all the coefficients $\Phi_{i,j}$, $A_{i,j}$ are uniquely determined from it. Note that $\zeta\neq\theta-2$ and $\psi\neq2$ for this solution to exist. The Chern-Simons coupling is responsible for generating the two modes seen in the expansion. The first stems from electric terms in the field equations, while the second from magnetic terms. The structure of the expansion is quite complicated, since these two types of terms mix already at quadratic order, but their amplitudes can be solved for consistently. They collapse to the same value for $\psi=\theta-\zeta$, or equivalently $2\lambda=\gamma_1+\gamma_2$ as for the nonrelativistic solutions.

The parameter space from the background and requiring the modes to be irrelevant is 
\be
\theta<0\,,\quad \xi<0\,,\quad \zeta<\theta-2\,,\quad \psi>2\,.
\ee

\section{AC conductivity\label{section:AC}}

\subsection{Massive vectors}

We now focus back on the theories \eqref{EMDaction2}, which we recall here for convenience
\be
\label{EMDaction5}
S=\int \ud^{d+2}x\sqrt{-g}\left[\mathcal R-\frac{\partial\phi^2}2-\frac{Z_0 e^{\gamma\phi}}4F^2-\frac{W_0 e^{\varepsilon\phi}}2A^2+V_0 e^{-\delta\phi}\right].
\ee
These theories have been argued to capture the low-temperature asymptotics of holographic superconductors in the presence of a logarithmically running condensate in the IR, \cite{gk2012}. This is to be contrasted with the cases examined in \cite{Gubser:2009, Horowitz:2009} which focussed on AS$_4$ or Lifshitz IR fixed points.

Beyond numerical calculations, the low-frequency behaviour of the AC conductivity can be extracted by reducing the fluctuation problem to a one-dimensional Schr\"odinger equation and matching the asymptotics in the IR and the UV, \cite{Horowitz:2009}. This technique was applied to the IR of holographic superconductors with a constant scalar in \cite{Horowitz:2009}, and to the IR of normal phases described by a Lifshitz \cite{KT} or hyperscaling violating \cite{cgkkm} geometry supported by a running scalar.

Here we shall compute the low-frequency behaviour of the AC conductivity for the hypercaling violating IR asymptotics of holographic superconductors. We start by introducing the vectorial perturbations at zero spatial momentum $a_x(t,r)$ and $g_{tx}(t,r)$ along one of the spatial directions $x\in R_{(d)}$ in the electric potential and metric
\be
\begin{split}
&\ud s^2=-D(r)\ud t^2+2g_{tx}(t,r)\ud t\ud x+B(r)\ud r^2+C(r)\ud R^2_{(d)}\\
&\phi=\phi(r)\,,\qquad A=A_t(r)\ud t+A_x(t,r)\ud x
\end{split}
\ee
Making use of time translation invariance, we can parameterize the time dependence of the fluctuations as $A_x(t,r)=a_x(r) e^{-i\omega t}$ and $g_{tx}(t,r)=g_{tx}(r)e^{-i\omega t}$. Plugging this Ansatz in, we obtain three non-trivial equations for the perturbations from the $x$ component of Maxwell's equations and the $tx$ and $rx$ components of Einstein's equations. The latter can be obtained from the first two. The $tx$ equation is a first-order equation for $g_{tx}(r)$, from which we can eliminate it completely from the non-trivial Maxwell's equation and obtain:
\be
 \partial_r\left(ZC^{\frac{d-2}2}\sqrt{D\over B} a_x'\right) +{ZC^{\frac{d-2}2}\left[
 \sqrt{B\over D} \omega^2 - \frac{(A_t')^2Z }{\sqrt{DB}}\right]}a_x-W\sqrt{DB}C^{\frac{d-2}{2}}a_x=0\,.
 \label{30a}\ee
Primes denote derivatives with respect to $r$.
Note that the background electric field $A_t'$ contributes an explicit term to the potential multiplying the term linear in $a_x$, as well as the mass $W$. This is a similar equation to that obtained in \cite{HoloSc}.

Changing variables to
\be
a_x=\frac{\tilde a}{\sqrt{\tilde Z}}\,,\qquad \tilde Z=Z C^{\frac{d-2}{2}}
\ee
as well as defining the Schr\"odinger coordinate
\be
\frac{\ud\tilde r}{\ud r}=\sqrt{\frac{B(r)}{D(r)}}
\ee
the fluctuation equation \eqref{30a} takes the form of a Schr\"odinger equation
\be\label{ACSchr}
-\frac{d^2\tilde a}{\ud\tilde r^2}+\tilde V\tilde a=\omega^2\tilde a
\ee
with
\be\label{SchrPotential}
\tilde V=\frac{(A_t')^2Z }{B}+\frac{W}{\tilde Z}DC^{\frac{d-2}{2}}+\frac14\frac{(\partial_{\tilde r}\tilde Z)^2}{\tilde Z^2}+\frac12\partial^2_{\tilde r}(\ln\tilde Z)
\ee
Note that for extremal, fractionalised backgrounds with $\epsilon\neq\gamma-\delta$, the mass term will actually be subleading in the IR and will not contribute to the Schr\"odinger potential. Moreover, the mass term is independent from the presence of a background electric field $A_t$, contrarily to the flux term proportional to $(A_t')^2$.

Assuming the hyperscaling violating form \eqref{HVLif}, then we find that $\tilde r\sim r^z$. In order to determine the locus of the extremal horizon, we have to consider where it stands in the $r$ coordinate. If we define it as the locus where the spatial part of the metric collapses, it is $r\to0$ if $\theta<d$ and $r\to\infty$ otherwise. Then the location of the horizon in the $\tilde r$ coordinate is decided by the sign of $(d-\theta)z$, that is by thermodynamic stability of the non-extremal black hole solution! For thermodynamically stable solutions, the extremal horizon will always be at $\tilde r\to+\infty$, while it will be at $\tilde r\to0$ otherwise. Note that this holds both for the $z=1$ and $z\neq1$ backgrounds. Let us restrict ourselves to the thermodynamically stable case, and assume the extremal horizon is at $r\to+\infty$.

Evaluating \eqref{SchrPotential} on the $z\neq1$ solutions \eqref{MassiveEMDsol} or \eqref{EMDsolCoh1}, we find that it always takes the form
\be
\tilde V(\tilde r)=\frac{\tilde V_0}{\tilde r^2}
\ee
which allows to solve the Schr\"odinger equation \eqref{ACSchr}. The value of $\tilde V_0$ will depend on the extremal background. In any case, \eqref{ACSchr} can be solved imposing ingoing wave boundary conditions at the extremal horizon such that
\be
\tilde a \sim e^{i\omega \tilde r},\qquad \tilde r\to \infty\,,
\ee
which yields the Hankel function
\be
\tilde a\sim \sqrt{\omega \tilde r}H^{(1)}_\nu(\omega\tilde r)\,,\qquad \nu=\sqrt{\tilde V_0+\frac14}\,.
\ee
The next step is to notice that \eqref{ACSchr} has a conserved flux
\be
\mathcal F=i\tilde a^\star \overset{\longleftrightarrow}{\partial_{\tilde r}}\tilde a\,,\qquad \partial_{\tilde r}\mathcal F
\ee
so that it can be computed both near the extremal horizon at $\tilde r\to-\infty$ and at the UV boundary $\tilde r\to0$. Using the matched asymptotic procedure of \cite{KT}, we finally derive that
\be
Re(\sigma)\sim\omega^{2\nu-1},\qquad T\ll\omega\ll\mu
\ee
where $\mu$ and $T$ are the chemical potential and temperature of the dual field theory.

Applying this procedure to the solutions \eqref{MassiveEMDsol}, we obtain for the IR Schr\"odinger potential
\be
\tilde V_0=\frac{(d-\theta-2+2z)(d-2-\theta+4z)}{4z^2}
\ee
and then for the AC conductivity scaling\footnote{Remember that there is also a delta function $\delta(\omega)$, which comes from the $1/\omega$ pole in the imaginary part of the AC conductivity, though it has different interpretations in the normal and superfluid phases, \cite{HoloSc}.}
\be\label{ACznot1}
Re(\sigma)\sim\omega^{\left|3-\frac2z+\frac{d_\theta}{z}\right|-1}.
\ee

We can also apply the calculation to phases where there is no background electric field at leading order, with $z=1$, such as \eqref{EMDsolCoh1}. Plugging in the values, we obtain
\be
\tilde V_0=\frac{\zeta(\zeta-2)}{4}
\ee
and then
\be\label{ACz1}
Re(\sigma)\sim\omega^{\left|1-\zeta\right|-1}.
\ee

Remarkably, both \eqref{ACznot1} and \eqref{ACz1} are identical to the results obtained in the massless case $W_0=0$ in \cite{cgkkm,gk}: this means that the leading scaling behaviour of the AC conductivity at low frequencies is the same in terms of these scaling exponents. This result also applies to the fractionalised solutions with $\epsilon\neq\gamma-\delta$. Of course, if one were to trade $z$ and $\theta$ for their expressions in terms of $\gamma$, $\delta$ and $W_0/V_0Z_0$, then the scalings would be different, and this would obscure the universality of the results above.

\subsection{Electron stars}

We can perturb the Einsten's equations coupled to the charged fluid of fermions with time-dependent perturbations
\be
A_x(t,r)=\frac{e}ka_x(r) e^{-i\omega t}\,,\qquad g_{tx}=g_{tx}(r) e^{-i\omega t}\,,\qquad u_x(t,r)=u_x(r) e^{-i\omega t}
\ee
and after some massaging, obtain the linearised equations
\be
\begin{split}
&g_{tx}'(r)+\frac{C'}{C}g_{tx}-2e^{\gamma\phi}h'a_x=0\\
& u_x(r)\left(\hat\rho+\hat p\right)+\hat\sigma a_x(r)=0\\
&a_x''+\left(\gamma\phi'+\frac{D'}{2D}-\frac{B'}{2B}\right)a_x'+\frac{B}{D}\omega^2a_x+Be^{(\epsilon-\gamma)\phi}\hat\sigma u_x+\frac{h'}{D}\left(g_{tx'}+\frac{C'}C g_{tx}\right)=0
\end{split}
\ee
which generalise previous results in \cite{ElectronStar}.
We can substitute for $g_{tx}$ and $u_x$ in the second-order differential equation for $a_x$:
\be
\frac{D}Ba_x''+\left(\gamma\phi'+\frac{D'}{2D}-\frac{B'}{2B}\right)\frac{D}Ba_x+\left(\omega^2-2e^{\gamma\phi}\frac{h'^2}{B}-\frac{D\hat\sigma^2}{\hat\rho+\hat p}e^{(\epsilon-\gamma)\phi}\right)a_x=0
\ee
 This equation can be reformulated as a one-dimensional Schr\"odinger equation using the same variables as in the previous section, with the potential
\be\label{ACSchrEstar}
\tilde V=\frac{(h')^2Z }{B}+\frac{D\hat\sigma^2}{(\hat\rho+\hat p)}e^{(\epsilon-\gamma)\phi}+\frac14\frac{(\partial_{\tilde r}\tilde Z)^2}{\tilde Z^2}+\frac12\partial^2_{\tilde r}(\ln\tilde Z)\,.
\ee
Remarkably, evaluating it on the solution \eqref{EstarCoh}
\be
\tilde V(\tilde r)={\tilde V_0}{\tilde r^2}=\frac{(\theta-2z)(\theta-4z)}{4z^2\tilde r^2}
\ee
which is identical to the result obtained with the massive vector backgrounds \eqref{MassiveEMDsol}. Accordingly, the AC conductivity also reads:
\be
Re(\sigma)\sim\omega^{\left|3-\frac{\theta}{z}\right|-1}.
\ee
Turning to the backgrounds \eqref{EstarCohz=1}, we find that the Schr\"odinger potential in \eqref{ACSchrEstar} evaluates to
\be
\tilde V(\tilde r)=2(\zeta-\xi-1)^2Q^2 \tilde r^{\zeta-\theta}+\frac{\zeta(\zeta-2)}{4\tilde r^2}
\ee
The flux term is subleading (since $\zeta\neq\theta-2$ for this solution), but not the density term. Carrying out the previous steps again, we find in the end that the AC conductivity scales like
\be
Re(\sigma)\sim\omega^{\left|1-\zeta\right|-1}
\ee
which once more is identical to the scaling\eqref{ACz1} for the massive vector field solutions.

\subsection{Discussion}

Let us now recall and discuss the two formul\ae\ we have obtained for the low-frequency scaling of the optical conductivity, which are valid for $T\ll\omega\ll\mu$:
\be\label{ResultACabsValue}
\begin{split}
z\neq1\,,\,\zeta=-d_\theta\,: &\qquad Re(\sigma)\sim\omega^{\left|3-\frac2z+\frac{d_\theta}z\right|-1}\,,\\
z=1\,,\,\zeta\neq-d_\theta\,: &\qquad Re(\sigma)\sim\omega^{\left|1-\zeta\right|-1}\,.
\end{split}
\ee
The first one holds for charged backgrounds with a relevant current and $z\neq1$, while in the second, the current is irrelevant and $z=1$. Note that we are always considering a translation-invariant, charged background, so that there is a delta function at zero frequency. Yet, for the $z=1$ case, note that the scaling obtained is identical to that of probe charge carriers in a neutral medium, which would not see any delta function. In this respect, there is a difference between irrelevant currents and probe charge carriers.

The result above, as we have seen, holds for very different systems, which may describe the IR of holographic superfluids or electron stars. In terms of the scaling exponents, it is universal. Trading the scaling exponents for parameters of the Lagrangian would only obscure that universality.

For $z\neq1$, if one imposes the Null Energy Condition as well as the thermodynamic stability condition, then the absolute value in the exponent is always positive,\footnote{Relaxing the latter while imposing the irrelevant deformation conditions only allows for negative values for the electron stars, which seems pathological.} which simplifies the expression to
\be\label{ACznot1Simpl}
\textrm{NEC+Thermo stability:}\quad Re(\sigma)\sim\omega^{2-\frac2z+\frac{d_\theta}{z}}\,,\qquad 2-\frac2z+\frac{d_\theta}{z}>0\,.
\ee
Then, the power exponent of the optical conductivity is always positive as well, which is reassuring since it is a linear response transport coefficient. It reduces to the correct result both for AdS$_2\times\mathbf R^d$ ($z\to+\infty$) or various Lifshitz solutions with $\theta=0$ (whether fractionalised or not). The optical conductivity scales like in a Lifshitz theory in $d_\theta$ dimensions.  This interpretation is also supported by the classes of hyperscaling violating solutions which can be lifted to appropriate scale invariant solutions, see \cite{gk, gk2012}.

Turning to the $z=1$ result, we can also impose the NEC as well as thermodynamic stability. On top of this we should add constraints that the current indeed behaves as a perturbation in the IR:
\be
(\zeta+d_\theta)d_\theta \leq0\,,\qquad \xi d_\theta\geq0\,.
\ee
This implies $1-\zeta>0$, so the $z=1$ AC conductivity scaling of \eqref{ResultACabsValue} simplifies to
\be
 Re(\sigma)\sim\omega^{-\zeta}\,,\qquad \zeta<\theta-d<0\,.
\ee
The optical conductivity always scales with a positive power, and setting $\zeta=\theta-d$ recovers the $z=1$ limit of \eqref{ACznot1Simpl}.

Both for $z=1$ and $z\neq1$, the absolute values in \eqref{ResultACabsValue} can be dropped. 
It is very tempting to write down the generic formula
\be
z\neq1\,,\,\zeta\neq-d_\theta\,:\qquad Re(\sigma)\sim\omega^{2-\frac2z-\frac{\zeta}z}\,,
\ee
which clearly reduces to the correct result for $z=1$. For this reason, we call $\zeta$ the \emph{conduction} exponent, since it seems to control the scaling of the optical conductivity whether $z=1$ or not. It would be very interesting to find phases where the value taken by $\zeta$ can be decorrelated from that taken by $z$.

The nature of the spectrum of fluctuations (gapped or gapless) can be determined by looking at the behaviour of the Schr\"odinger potential in the IR. Note that the Schr\"odinger coordinate
\be
\tilde r\sim r^{z}
\ee
which means that for thermodynamically stable phases,
\be
\frac{d_\theta}z>0\,,
\ee
and the IR is always to $\tilde r\to+\infty$. In that case, the Schr\"odinger potential vanishes in the IR, and consequently the spectrum is gapless.
\begin{figure}
\begin{tabular}{cc}
\includegraphics[width=.45\textwidth]{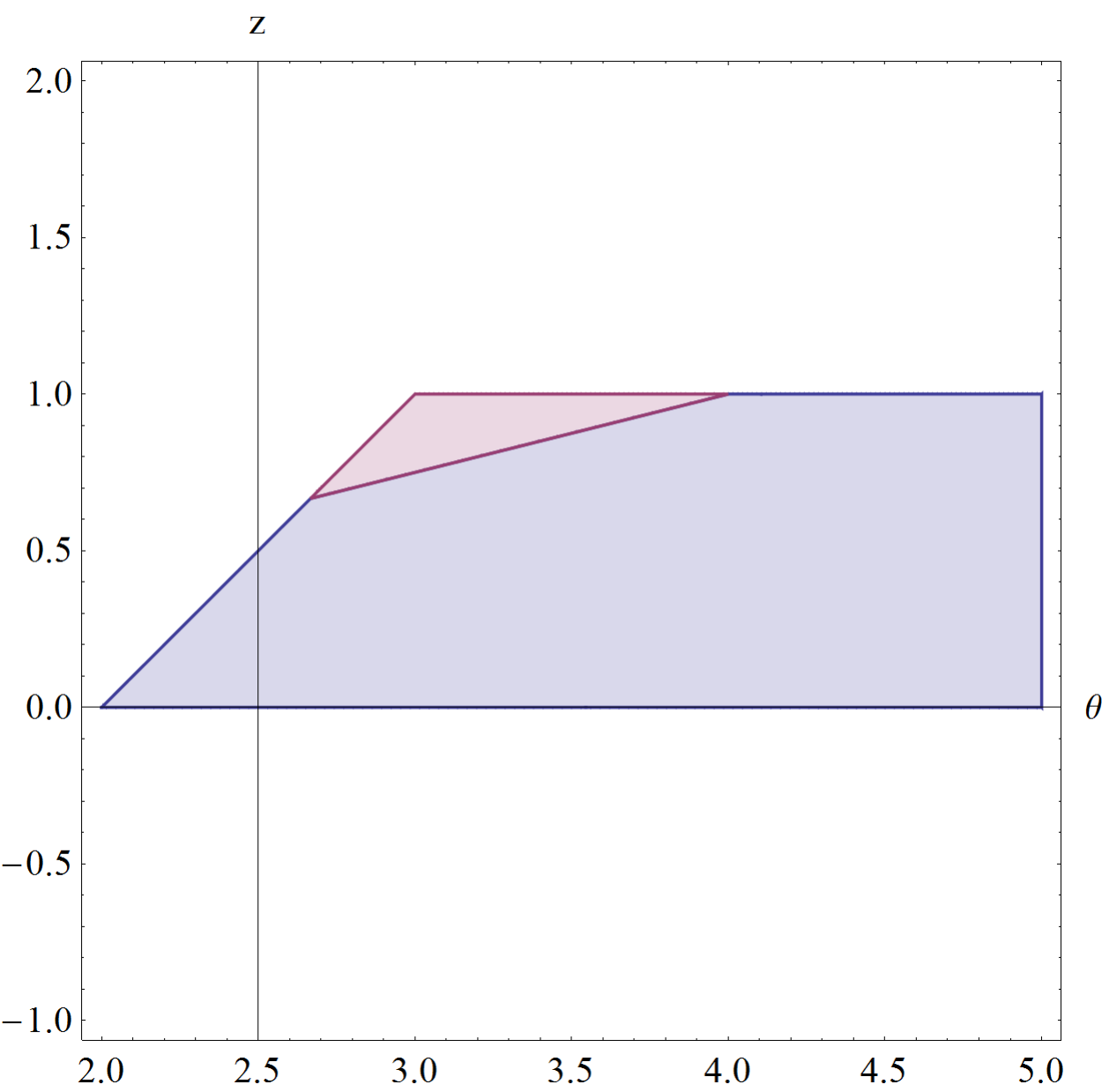}&\includegraphics[width=.45\textwidth]{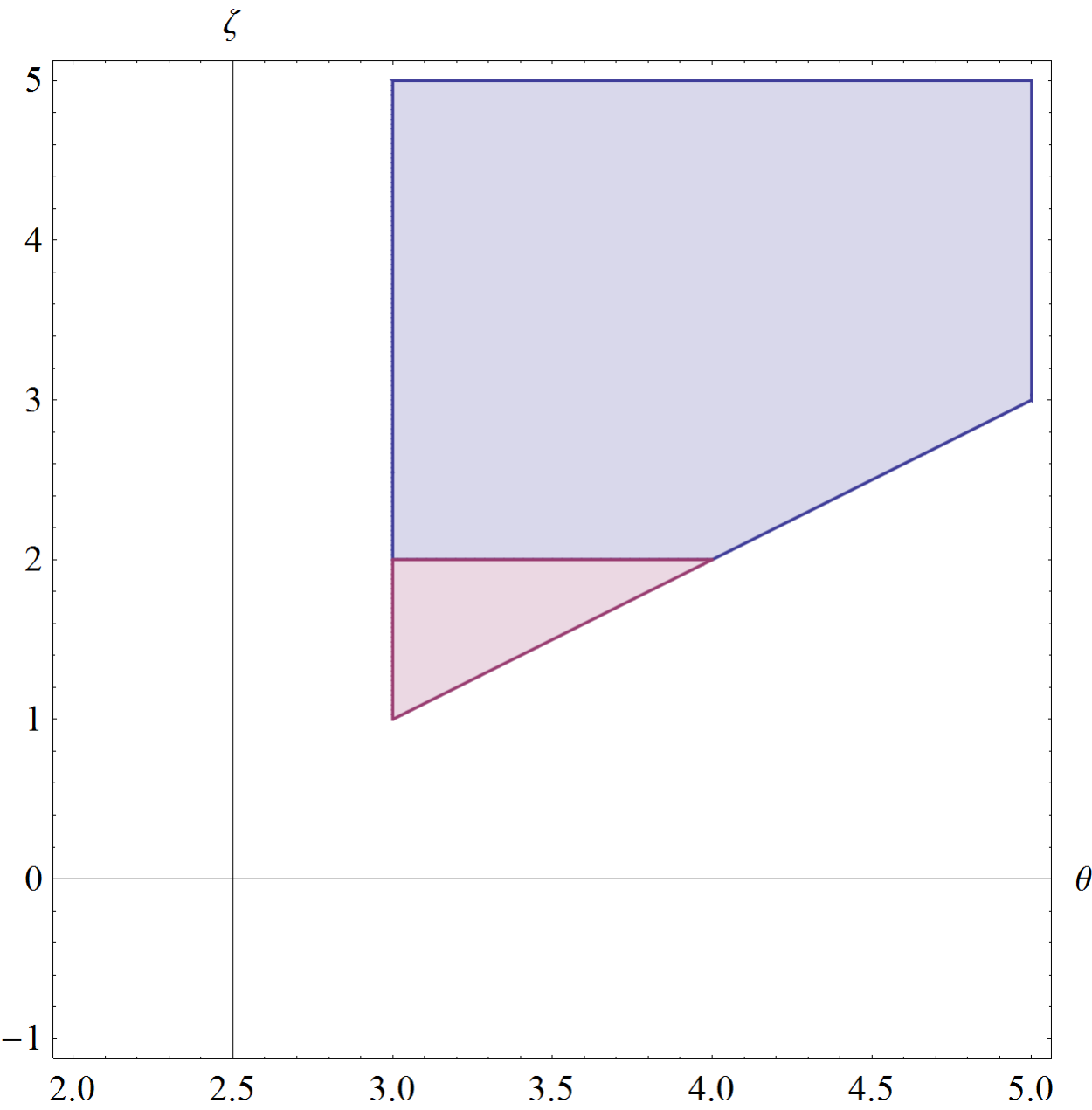}
\end{tabular}
\caption{Behaviour of the Schr\"odinger potential in the IR in $d=2$ (Left pannel: $z\neq1$; right pannel: $z=1$). In the red region, it diverges to $-\infty$, the optical conductivity vanishes at low frequency and the spectrum is gapless. In the blue region, it diverges to $+\infty$, linear response theory breaks down and the spectrum will be gapped if the potential also diverges in the UV.}
\label{fig:spectrum}
\end{figure}

In the thermodynamically unstable region (see figure \ref{fig:spectrum} for $d=2$), the Schr\"odinger potential blows up in the IR, and the nature of the spectrum can differ whether to $-\infty$ ($\tilde V_0<0$) or $+\infty$ ($\tilde V_0>0$).
If $\tilde V_0<0$, then the optical conductivity also vanishes at $\omega\to0$, so the spectrum is still gapless. 
If $\tilde V_0>0$, the optical conductivity blows up at $\omega\to0$, linear response theory breaks down and our result is not valid. If on top of this the Schr\"odinger potential also blows up in the UV,\footnote{This is always the case in $d>2$, and happens in $d=2$ if the scalar operator has a UV conformal dimension $1/2<\Delta<1$, \cite{cgkkm}.} then the spectrum will be gapped. More generally, in this case, we expect that there will be a phase transition to another branch of black holes at low enough temperature, so that this IR is not reached anyway.

\section{Revisiting the deformed entropy proposal\label{section3}}

In section \ref{section1}, we have shown how cohesive phases can be universally parameterized using four scaling exponents. Two are familiar: the dynamical exponent $z$, which measures the anisotropy between time and space of the dual field theory, and the hyperscaling violation exponent $\theta$, which tells us the effective spatial dimensionality;
 the third is the \emph{cohesion} exponent $\xi$ and measures the amount of electric flux in the IR, while the fourth, the \emph{conduction} exponent $\zeta$ controls the low-frequency scaling of the optical conductivity. Together, $\xi$ and $\zeta$ also control how much the electric potential departs from invariance under Lifshitz scaling. This parameterization holds in a variety of models, for  fractionalised as well as cohesive phases, independently from the details of the bulk source for electric charge. Let us recall this parameterization for convenience:
\be\label{LifLike0}
\begin{split}
	&\ud s^2=r^{\frac2d\theta}\left(\frac{L^2\ud r^2+\ud x^2_{(d)}}{r^2}- \frac{\ud t^2}{r^{2z}}\right),\\
	&A=Q r^{\zeta -\xi-z}\ud t\,,\qquad \int Z(\phi)\star F\sim r^\xi\,.
\end{split}
\ee
While the interpretation of these scaling exponents as measuring the departure from scale invariance
\be\label{LifScaling2}
r\to\lambda r\,,\qquad x\to\lambda x\,,\qquad t\to\lambda^z t
\ee
 is most obvious in this system of coordinates, it will prove useful in the following to introduce a coordinate $\rho=r^{\frac{d}{d_\theta}}$ for which the IR is unambiguously $\rho\to+\infty$:
\be\label{LifLike}
\begin{split}
	&\ud s^2=\rho^{-2}\ud x^2_{(d)}+\tilde L^2\rho^{2\frac{2\theta-d}{d_\theta}}\ud \rho^2 -\rho^{2\frac{\theta-dz}{d_\theta}}\ud t^2\,,\\
	&A= Q \rho^{\frac{d(\zeta-\xi-z)}{d_\theta}}\ud t\,,\qquad \int Z(\phi)\star F\sim \rho^{\frac{d\xi}{d_\theta}}\,.
\end{split}
\ee
Beyond the interpretation we just mentioned, we have seen in section \ref{section:AC} how $\zeta$ controls the scaling of the optical conductivity. How about the cohesion exponent $\xi$?

Since we are looking for observables characterising the IR phase without making any reference to the UV completion, thermodynamic quantities are of little use.
One might like to think about entanglement entropy, which is a nonlocal probe of entanglement between a spatial subsystem of the boundary. Holographically, it is calculated by minimizing the area of the bulk hypersurface whose boundary is the boundary of the subsystem, \cite{RyuTaka}. If the typical size of the boundary subsystem is large enough, then the details of the UV can be neglected and a universal IR piece extracted, \cite{Ogawa:2011bz}. However, it is only sensitive to the spatial part of the metric, and so 'sees' neither $z$ nor $\zeta$ or $\xi$.

To remedy this, we will take up the analysis of \cite{Hartnoll:2012ux}, which proposed a modification of the Ryu-Takayanagi formula so that it might be sensitive to the presence of cohesive/fractionalised charges in the IR. To be more precise, these authors suggested two observables:
\begin{itemize}
\item calculate the electric flux through the minimal surface $\Gamma$ determined by the calculation of the entanglement entropy;
\item minimize a 'deformed' entanglement entropy 
\be
S_E^\lambda=\frac{A_\Gamma}{4 G_N}+\lambda\Phi_\Gamma\,.
\ee
\end{itemize}

These authors addressed the case of a constant charge density in the IR, and found that the physics of these observables could be classified according to the sign of $\theta$. Our discussion will parallel that of \cite{Hartnoll:2012ux}, and we will show that (perhaps unsurprisingly), it is the cohesion exponent $\xi$ which controls the physics and determines whether the minimal surface found differs from the Ryu-Takayanagi prescription or not. The fact that for Lifshitz-like asymptotics \eqref{LifLike} a constant charge density is recovered by setting $\xi=\theta-d+\theta/d$ ties back to the results of \cite{Hartnoll:2012ux}.

Motivated by the results of section \ref{section1}, we will generalize this discussion to include a density scaling in the IR $\rho\to\infty$:
\be\label{sigma2}
\sigma=\sigma_0 \rho^{\sigma_1}\,, \qquad \sigma_1=\frac{d}{d_\theta}(d_\theta -\frac{\theta }{d}+\xi)
\ee
while the electric flux reads:
\be\label{ElFlux2}
E(\rho)= Q \rho^{\frac{d\xi}{d_\theta }}\,.
\ee
These expressions relate to those of the previous sections by the change of radial coordinate $\rho=r^{d_\theta/d}$. Anticipating on what follows, we will require $d_\theta\geq 0$, so that the flux will vanish in the IR ($\rho\to+\infty$) if $\xi\leq0$. As previously, $\xi=0$ recovers a constant flux and a fractionalised phase.

As we will closely follow the procedure of \cite{Hartnoll:2012ux}, we will only streamline the discussion and refer to this work for further details.

We are now in position to calculate what is the minimal spatial hypersurface using the metric \eqref{LifLike}. For simplicity we will only perform the calculation for a strip of length $L$ along one of the spatial boundary directions, which extends uniformly along the other directions with volume $Vol(\Sigma)$. We have to minimize the area
\be\label{AreaMin}
A_\Gamma=Vol(\Sigma)\int\frac{\ud \rho}{\rho^d}\sqrt{\rho^{\frac{2\theta}{d_\theta}}+\dot x^2}\,.
\ee
The surface $\Gamma$ minimizing this area is found to verify
\be \label{Xdot}
\dot x^2 = \frac{\rho^{\frac{2\theta}{d_\theta}}}{\left(\rho_0/\rho\right)^{2d}-1}
\ee
where $\rho_0$, the locus where $\dot x$ diverges, is the maximal radius reached by the surface. Let us now calculate the separation $L$ in $x$ on the boundary:
\be
L=\int \ud x = \int \ud \rho\, \dot x = \int_0^{\rho_0} \ud \rho\frac{\rho^{\frac{\theta}{d_\theta}}}{ \sqrt{\left(\rho_0/\rho\right)^{2d}-1}}\sim \rho_0^{\frac{d}{d_\theta}}
\ee
assuming that the we can approximate the integral in the IR close to $\rho_0$ with the IR geometry \eqref{LifLike}. As we will require $d_\theta\geq 0$ (so that the entanglement entropy cannot grow faster than a volume law), this means that $L$ grows when $\rho_0$ does, so it is indeed the dominant piece in the entanglement entropy in the IR region and we can forget about the contribution coming from the $0\leq\rho\ll\rho_0$ piece of the integral. That part of the integral simply contributes a non-universal constant independent from $L$, which results into a boundary law for the entanglement entropy at large $L$.

Next, we can compute the area of $\Gamma$ using \eqref{AreaMin} and \eqref{Xdot}:
\be\label{Agamma}
A_\Gamma\sim Vol(\Sigma)L^{1-d_\theta}
\ee
which grows at most like a volume law if $d_\theta\geq 0$.

Let us now compute the electric flux threading the minimal surface $\Gamma$:
\be\label{PhiGamma1}
\begin{split}
\Phi_\Gamma &= Vol(\Sigma)\int E(\rho)\ud x =  Vol(\Sigma)\int \ud \rho \rho^{\frac{d\xi}{d_\theta }}\frac{\rho^{\frac{\theta}{d_\theta}}}{\sqrt{\left(\rho_0/\rho\right)^{2d}-1}}\\
		&= Vol(\Sigma)\rho_0^{\frac{d(1+\xi)}{d_\theta}}\sim Vol(\Sigma) L^{1+\xi}
\end{split}
\ee
Since we do not want the flux to diverge in the IR, we have imposed the constraint $\xi\leq0$. Then, $\Phi_\Gamma$ always grows less than a volume law, which is reached for the (partially) fractionalised case $\xi=0$. For a constant density, we get $\xi=-d_\theta +\theta/d$, this reduces to the case examined by \cite{Hartnoll:2012ux}. More generically, $\xi<0$ and \eqref{PhiGamma1} will be give the scaling for the cohesive geometries we examined in section \ref{section:MassEMDsol}.

Whenever $\xi<-d_\theta$, $\Phi_\Gamma$ will be subleading compared to the entanglement entropy $A_\Gamma$ in \eqref{Agamma}, while the reverse holds for $\xi>-d_\theta$.
If $\xi=-d_\theta$, both terms scale identically. For phases with $z\neq1$, this means $\xi=\zeta$ and the electric potential is invariant under the Lifshitz scaling \eqref{LifScaling2}.

Let us now turn to the minimization of the deformed entanglement entropy itself:
\be
S^\lambda_E=\frac{A_\Gamma}{4G_N}+\lambda\Phi_\Gamma=Vol(\Sigma)\int\ud \rho\,\left(\frac{1}{4G_N\rho^d}\sqrt{\rho^{\frac{2\theta}{d_\theta}}+\dot x^2}+\lambda\sigma_0\dot x \rho^{\frac{d \xi}{d_\theta }}\right)
\ee
Minimizing the area above, we obtain the following constraint on $\dot x$:
\be\label{D(rho)1}
\dot x^2=\frac{\rho^{\frac{2\theta}{d_\theta}}D(\rho)^2}{\left(1/4G_N\right)^2-D(\rho)^2}\,, \qquad D(\rho)=C\rho^d-\lambda\sigma_0 \rho^{d+\frac{d\xi}{d_\theta}}\,,
\ee
which gives the profile of the minimal surface $\Gamma$. The integration constant $C$ comes from the fact that the area only depends explicitly on $\dot x$, so that the Euler-Lagrange equations integrate straightforwardly. The maximal radius $\rho_0$ reached by $\Gamma$ is given by the locus at which $\dot x$ diverges, $|D(\rho_0)|=1/4G_N$. Setting aside for now the question whether such a hypersurface, when it exists, actually extends all the way to the boundary, we now compute the IR contribution of the minimal surface supported by $\Gamma$ to the area $A_\Gamma$ and the electric flux $\Phi_\Gamma$. To do this, we assume that for $\rho_0$ large this contribution is well approximated by restricting the integration domain to $\rho_0(1-\epsilon)\leq \rho<\rho_0$. Expanding $D(\rho)=D(\rho_0)+(\rho-\rho_0)D'(\rho_0)$ around $\rho_0$ and using $D(\rho_0)^2=1/4G_N$, we find
\be
D'(\rho_0)\sim\left\{\begin{array}{cc}\rho_0^{-1},&\quad\xi\leq-d_\theta\\ \rho_0^{-1+d+\frac{d\xi}{d_\theta}}, &\quad\xi>-d_\theta\end{array}\right.
\ee
where we have kept the dominant contribution at large $\rho_0$, depending on the sign of $\xi-\theta+d$. For the geometries \eqref{MassiveEMDsol}, the two cases are distinguished once more by the value of $\xi$.

We can now determine the length of the minimal surface through
\be
L\sim\int_{\rho_0(1-\epsilon)}^{\rho_0}\ud \rho\,\dot x\sim\left\{\begin{array}{cc}\rho_0^{\frac{d}{d_\theta}},&\quad\xi\leq-d_\theta\\ \rho_0^{\frac{d(2-d_\theta-\xi)}{2d_\theta}}, &\quad\xi>-d_\theta\end{array}\right.
\ee
Note that the length $L$ grows with $\rho_0$ only if $\xi\leq2-d_\theta$. Otherwise, the details of the full spacetime are needed, since we cannot restrict the integrals to the IR contribution in the region close to $\rho_0$.
Next comes the contribution to the area $A_\Gamma$, which reads
\be
\frac{A_\Gamma}{Vol(\Sigma)}\sim\int_{\rho_0(1-\epsilon)}^{\rho_0}\ud \rho\,\frac1{\rho^d}\sqrt{\rho^{\frac{2\theta}{d_\theta}}+\dot x^2}\;\sim\left\{\begin{array}{cccc}\rho_0^{\frac{d}{d_\theta}-d}&\sim& L^{1-d_\theta},&\quad\xi\leq-d_\theta\\ \rho_0^{\frac{d(2 -d_\theta-\xi)}{2d_\theta}-d}&\sim& L^{\frac{2-\xi-3d_\theta}{2-d_\theta-\xi}}, &\quad\xi>-d_\theta\end{array}\right.,
\ee
while the contribution to the electric flux is
\be
\frac{\Phi_\Gamma}{Vol(\Sigma)}\sim\int_{\rho_0(1-\epsilon)}^{\rho_0}\ud \rho\,\dot x \rho^{\frac{d \xi}{d_\theta }}\;\sim\left\{\begin{array}{cccc}\rho_0^{\frac{d(1+\xi)}{d_\theta}}&\sim& L^{1+\xi},&\quad\xi\leq-d_\theta\\ \rho_0^{\frac{d(2-\zeta)}{2d_\theta}-d}&\sim& L^{\frac{2-d_\theta+\xi}{2 -d_\theta-\xi}}, &\quad\xi>-d_\theta\end{array}\right.,
\ee
Given $d_\theta>0$, it is clear that for $\xi\leq-d_\theta$, the area term dominates the deformed entanglement entropy at large $L$ and the minimal surface is the Ryu-Takayanagi one. The flux term dominates if $\xi>-d_\theta$ (remembering that the calculation is valid when $\xi\leq2-d_\theta$), and in this case the minimal surface differs. When the inequality is saturated, both contribute the same. For the geometries \eqref{MassiveEMDsol}, this means $\xi=\zeta=-d_\theta$, which intriguingly is the case where the gauge field is left invariant by the Lifshitz rescaling. For $\xi\leq-d_\theta$, the dominant (area) term is at most linear at large $L$, where it grows like a volume law. The fractionalised case $\xi=0$ occurs when the flux term dominates, and we recover a volume law $\Phi_\Gamma\sim Vol(\Sigma)L$, as expected. Still when $\xi<-d_\theta$, there is a range for which the flux term grows with $L$:
\be
-2<\xi <0\,,\qquad\,d_\theta-2 <\xi \leq d\,,\qquad d_\theta\leq0
\ee
with a growth bounded from above by the volume scaling of the fractionalised case.

\begin{figure}\label{Fig:D1}
\begin{tabular}{cc}
\includegraphics[width=.47\textwidth]{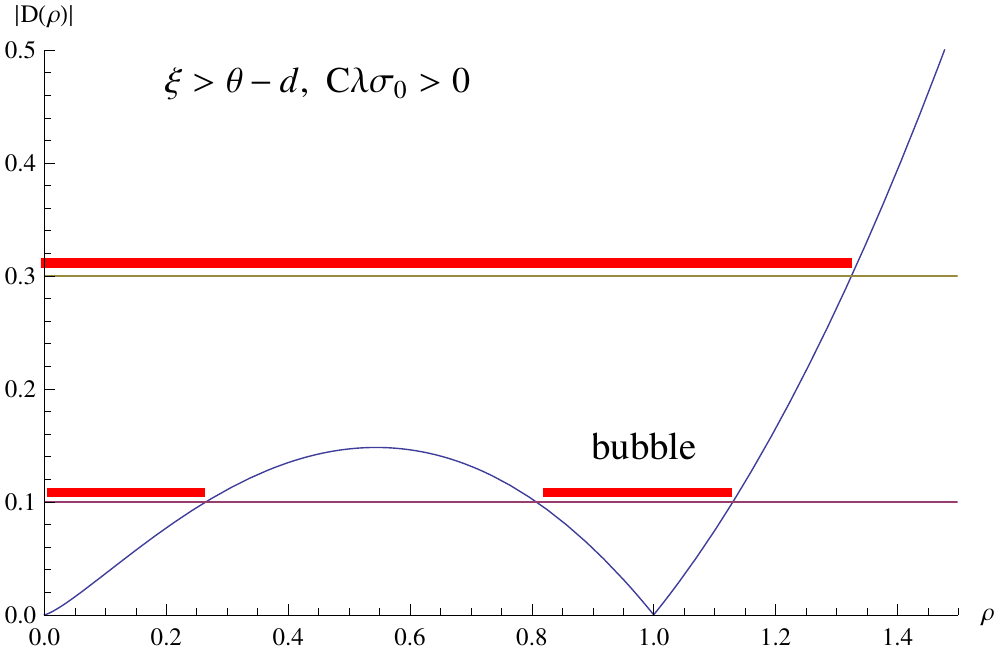}&\includegraphics[width=.47\textwidth]{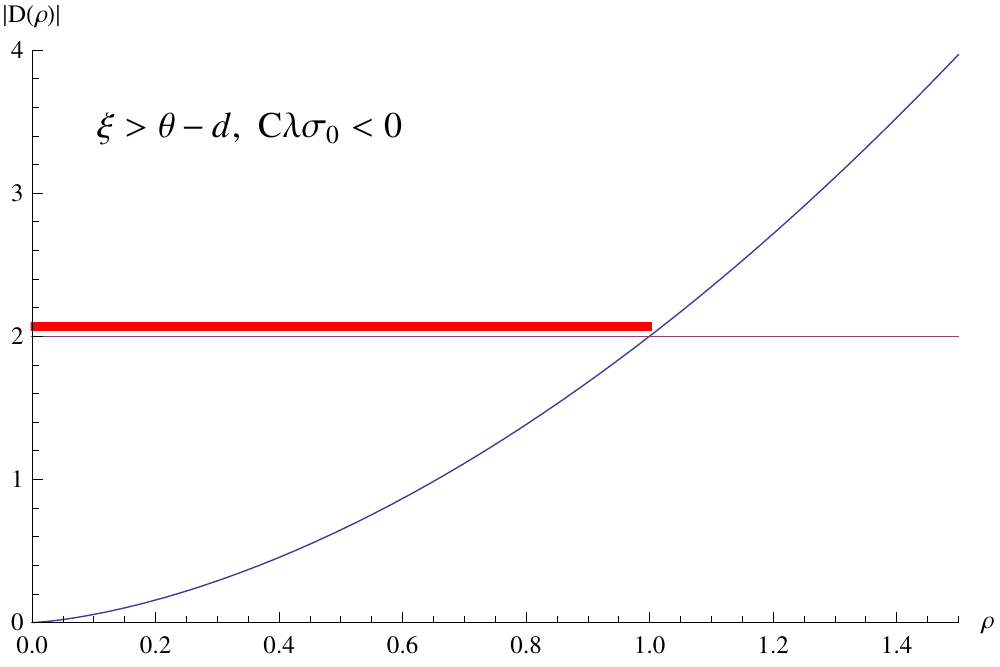}\\
\includegraphics[width=.47\textwidth]{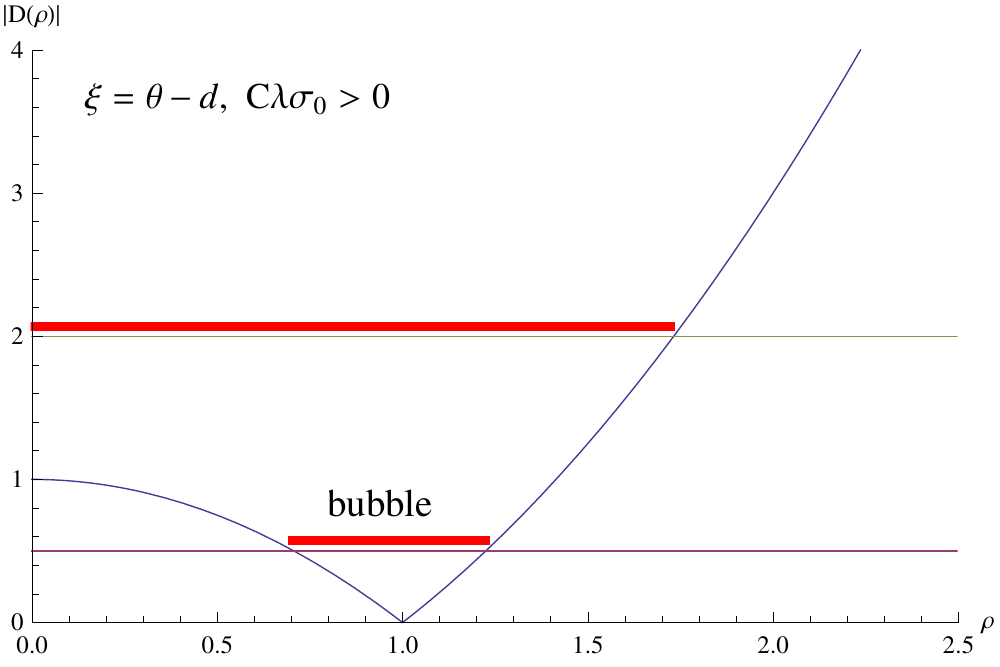}&\includegraphics[width=.47\textwidth]{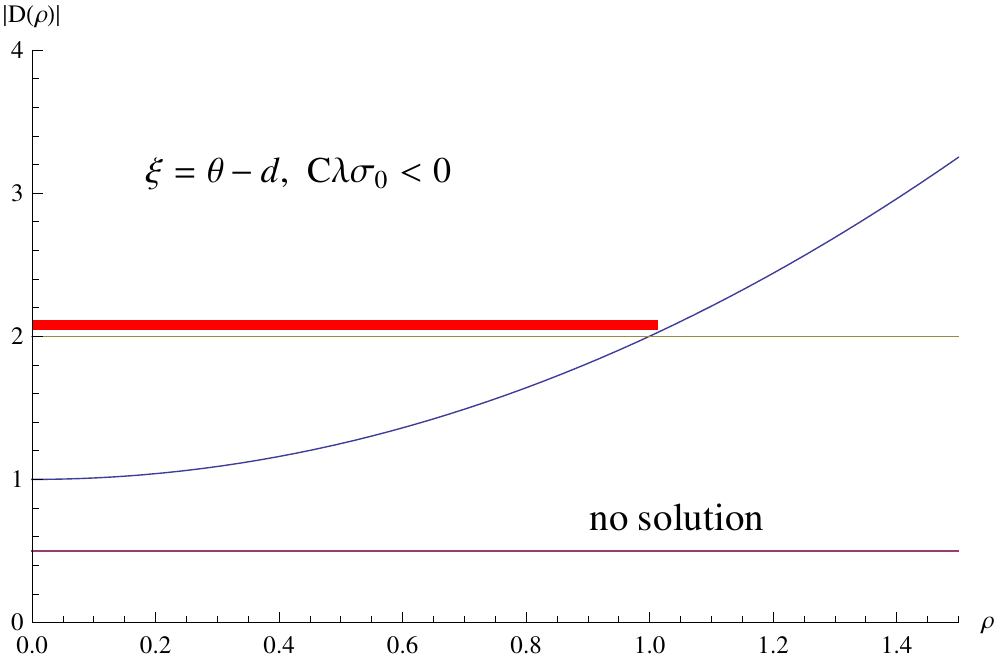}\\
\includegraphics[width=.47\textwidth]{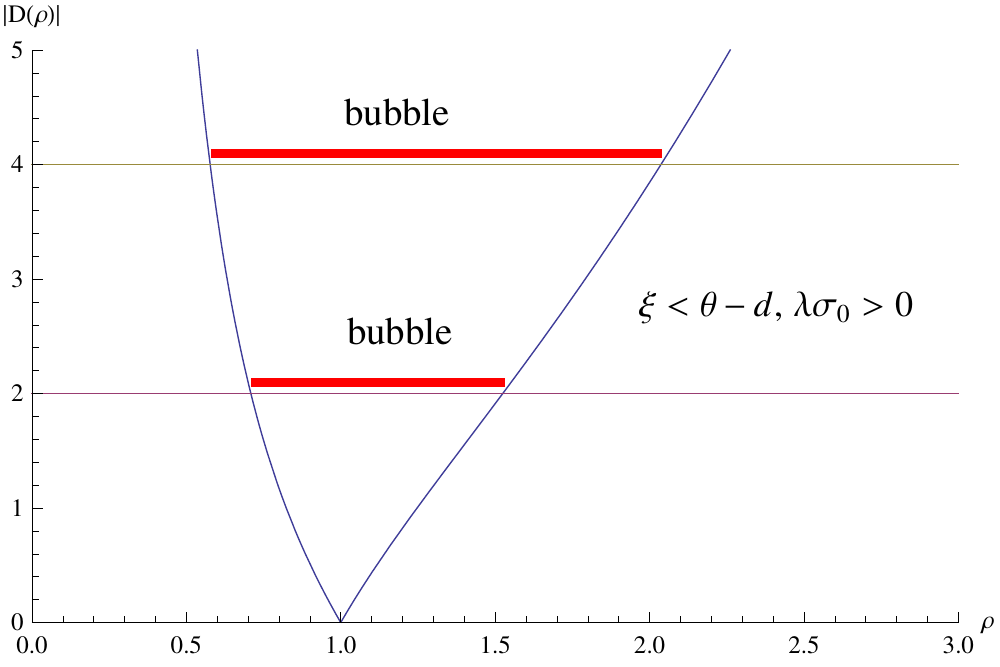}&\includegraphics[width=.47\textwidth]{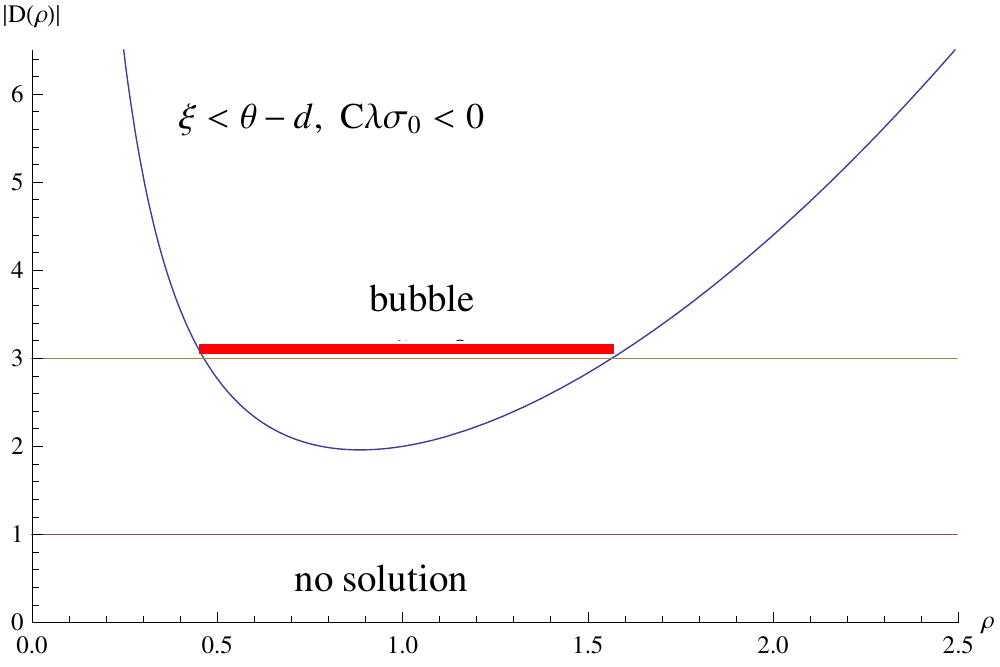}
\end{tabular}
\caption{Typical behaviour of the function $D(\rho)$ depending on the sign of $\xi$ and the relative signs of $\lambda\sigma_0$ and the integration constant $C$. The horizontal lines represent different values of $1/4G_N$, and solutions of $D(\rho_0)=1/4G_N$ are highlighted in red and with a thicker line. The solutions can either reach the boundary with or without a disconnected piece in the bulk (a bubble), or there might be no solution, or just a bubble in the bulk which does not reach the boundary (though see main text).}
\end{figure}

The last point to address is whether the hypersurfaces $\Gamma$ can actually exist for all values of $\xi$ and of the other parameters appearing in $D(\rho)$. Inspecting \eqref{D(rho)1}, it is easy to convince oneself that $D(\rho)\sim \rho^d$ as $\rho\to\infty$ while $D(\rho)\sim \rho^{d+d\xi/d_\theta}$ as $\rho\to0$, given the constraints $d_\theta>0$ and $\xi\leq0$ (which ensures that the electric flux \eqref{ElFlux2} does not blow up in the IR). The various cases are spelled out below and summarized in the plots of Fig. \ref{Fig:D1}.

If $\xi>-d_\theta$ (the flux term dominates $S_E^\lambda$), $D(\rho)$ vanishes as $\rho\to0$ while it blows up in the IR, guaranteeing that for values of $G_N$ such that $|D(\rho_0)|\leq 1/4G_N$, there will be a solution that reaches the boundary, possibly with a disconnected piece in the bulk (a bubble).

For $\xi=-d_\theta$ (the flux and area terms scale indentically in $S_E^\lambda$), the IR value of $D(\rho)$ will just be a constant, and only if $|\lambda\sigma_0|>1/4G_N$ will the hypersurface reach the boundary (in this case, the coefficient of the flux term is greater than the area term). 

Finally, if $\xi<-d_\theta$ (the area term dominates the deformed entanglement entropy), the function $|D(\rho)|$ will diverge on both ends of the coordinate range in $\rho$ and have a minimum inbetween: consequently, whether the hypersurface $\Gamma$ can reach the boundary in that case cannot be decided just by inspecting the IR geometry. The full interpolation to the UV AdS$_4$ is needed, and then we expect the behaviour of $D(\rho)$ as $\rho\to0$ to be regulated (remember that the $\rho$ coordinate does not go all the way to the boundary, simply to the UV of the IR region). It will acquire a maximum in the intermediary region inbetween the IR region and the boundary region, and the hypersurfaces $\Gamma$ will be well-defined for values of the parameters such that this barrier can be overcome. For values where this does not happen, then the minimal surfaces will behave rather like in confining geometries and not reach all the way to the IR.

It is interesting to note that this is the case where one might expect the deformed entanglement entropy to resemble most the RT prescription, since it is dominated by the area term. However, the presence of flux in the IR modifies substantially the dynamics of hypersurfaces minimizing the HR prescription, and hence comparison of the two prescriptions allows to distinguish whether the phase is cohesive or not.

\section{Outlook \label{section:conclusions}}

In this work, we have proposed a universal parameterization of IR cohesive phases (with vanishing electric flux). It relies on four scaling exponents: the dynamical exponent $z$ which measures anisotropicity between time and space; the hyperscaling violation exponent $\theta$ which measures the departure of the metric from scale invariance; the cohesion exponent $\xi$ which describes the scaling of the electric flux; and the conduction exponent $\zeta$ which, together with $\xi$, measures the departure of the electric potential from scale invariance. These exponents can be determined from data in the Lagrangian, and so give a hands-on recipe to determine which phases (maintaining translation invariance) might be competing in the IR. We have shown that this parameterization holds in a variety of models (effectively presenting large new families of solutions), where Gauss's law in the IR is broken either by the introduction of a bosonic condensate, a fermion fluid or a magnetic field. We have also commented on how these effective mass terms for the vector field need to be tuned so as to be relevant in the IR: otherwise, (partially) fractionalised phases are generated, with constant electric flux emanating from the charged extremal horizon.

Generically, solutions fall into two classes:
\begin{itemize}
\item Either they are nonrelativistic ($z\neq1$), and then the conduction exponent takes the value $\zeta=-d_\theta$;
\item or relativistic symmetry is restored in the IR ($z=1$) and then the conduction exponent is independent of the other scaling exponents.
\end{itemize}
An obvious extension of this formalism would be to consider phases which break translation invariance, perhaps starting along the lines of \cite{kasa}.

We have examined the (ir)relevant deformations around these phases and show that they always come by pairs summing to $d_\theta+z$, except if the current is irrelevant but cannot be switched off or coupled to a magnetic field via a Chern-Simons term: then, interestingly, there is an anomalous contribution controlled by $\zeta$. The precise values typically involve all critical exponents $z$, $\theta$, $\zeta$ and $\xi$.

We have then studied the optical conductivity at low frequencies and argued that its scaling is controlled by the conduction exponent $\zeta$. This is a universal result, which does not depend on the details of the Lagrangian, and holds both for cohesive and fractionalised phases. We have only examined the models without magnetic field. Upon turning on a magnetic field, one expects the relevant conductivity to be the Hall conductivity, which has been calculated in certain scale invariant cases in \cite{Hall}. The problem is made harder because various perturbations now do not decouple as straightforwardly as in the pure electric case, but we hope to return to this problem in the future. Furthermore, the spectrum of electric fluctuations is always gapless in the thermodynamically stable region.

Turning to nonlocal observables, we have examined a recent proposal of a deformed entanglement entropy, and showed that its scaling behaviour for large boundary subsystems is determined by the value of the cohesion exponent: for large enough values, the scaling found differs from the Ryu-Takayanagi prescription; for small values, even though the scaling is identical with the RT prescription, the dynamics of the minimal surface might be very different from the RT prescription. Combining these two observables thus gives a good probe of the presence of electric flux in the IR. Intriguingly, the cohesion exponent also knows about the violation of the Lifshitz scaling in the electric potential, since both contributions in the deformed entanglement entropy scale identically in the scale invariant case.

A number of works have recently addressed the reconstruction of bulk data from boundary data linked to entanglement entropy and extremal surfaces, \cite{RecBulk}. But as we already stressed, this seems oblivious to the presence of flux or not in the IR, which from a boundary point of view means that it does not distinguish between those horizon degrees of freedom charged under the boundary U(1) and those which are neutral. The idea (left for future work) would be to assert the validity of an equation like
\be
S_E^\lambda=\textrm{Tr}(\rho_{U(1)}\log\rho_{U(1)})
\ee
where $\rho_{U(1)}$ would now only be the density matrix associated to the degrees of freedom carrying the U(1) charge.

\section*{Acknowledgements}

I am grateful to R. Davison, A. Donos, D. Hofman and E. Kiritsis for very interesting discussions on this topic. I would especially like to thank S. Hartnoll and L. Huijse for extensive discussions on their work \cite{Hartnoll:2011pp}. I am also thankful to E. Kiritsis for comments on the manuscript. I would like to thank the organizers of the Banff workshop on 'Holography and applied String Theory' where this project first took shape, as well as CCTP (U. of Crete) for hospitality over the course of this work. It was also cofinanced by the European Union (European Social Fund, ESF) and Greek national funds
through the Operational Program Education and Lifelong Learning" of the National Strategic Reference
Framework (NSRF) under Funding of proposals that have received a positive evaluation in the 3rd and
4th Call of ERC Grant Schemes.


\addcontentsline{toc}{section}{References}
\begin{thebibliography}{99}


\bibitem{Hartnoll:2011fn}
  S.~A.~Hartnoll,
  {\em``Horizons, holography and condensed matter,''}
 in {\it Black Holes in Higher Dimensions},  Edited by: Gary T. Horowitz,
Cambridge University Press, 2012.
  \hri{1106.4324}{[hep-th]}.

\bibitem{Iqbal:2011ae}
  N.~Iqbal, H.~Liu and M.~Mezei,
  {\em``Lectures on holographic non-Fermi liquids and quantum phase transitions,''}
  \hri{1110.3814}{[hep-th]}.


 \bibitem{cgkkm}
  C.~Charmousis, B.~Gout\'eraux, B.~S.~Kim, E.~Kiritsis, R.~Meyer,
  {\em ``Effective Holographic Theories for low-temperature condensed matter systems,''}
  JHEP {\bf 1011 } (2010)  151.
 \hri{1005.4690}{[hep-th]}.

\bibitem{HoloSc}
  S.~A.~Hartnoll, C.~P.~Herzog and G.~T.~Horowitz,
{\em``Building a Holographic Superconductor,''}
  Phys.\ Rev.\ Lett.\  {\bf 101} (2008) 031601
  \hri{0803.3295}{[hep-th]};
\\
  {\em``Holographic Superconductors,''}
  JHEP {\bf 0812} (2008) 015
  \hri{0810.1563}{[hep-th]}.


  \bibitem{gk}
  B.~Gout\'eraux and E.~Kiritsis,
  {\em ``Generalized Holographic Quantum Criticality at Finite Density,''}
  JHEP {\bf 1112} (2011) 036
  \hri{1107.2116}{[hep-th]}.

\bibitem{gk2012}
  B.~Gout\'eraux and E.~Kiritsis,
  JHEP {\bf 1304} (2013) 053
  \hri{1212.2625}{[hep-th]}.


  \bibitem{sachdev}
  L.~Huijse, S.~Sachdev and B.~Swingle,
  {\em ``Hidden Fermi surfaces in compressible states of gauge-gravity duality,''}
  Phys.\ Rev.\ B {\bf 85} (2012) 035121
  \hri{1112.0573}{[cond-mat.str-el]}.


\bibitem{MagIRres}
  S.~Harrison, S.~Kachru and H.~Wang,
  {\em``Resolving Lifshitz Horizons,''}
  \hri{1202.6635}{[hep-th]}.
\\
J.~Bhattacharya, S.~Cremonini and A.~Sinkovics,
 {\em ``On the IR completion of geometries with hyperscaling violation,''}
JHEP {\bf 1302} (2013) 147
  \hri{1208.1752}{[hep-th]}.
\\
 N.~Kundu, P.~Narayan, N.~Sircar and S.~P.~Trivedi,
  {\em``Entangled Dilaton Dyons,''}
JHEP {\bf 1303} (2013) 155
  \hri{1208.2008}{[hep-th]}.


  \bibitem{Ogawa:2011bz}
  N.~Ogawa, T.~Takayanagi and T.~Ugajin,
  {\em  ``Holographic Fermi Surfaces and Entanglement Entropy,''}
  JHEP {\bf 1201} (2012) 125
   \hri{1111.1023}{[hep-th]}.



  \bibitem{dong}
  X.~Dong, S.~Harrison, S.~Kachru, G.~Torroba and H.~Wang,
  {\em ``Aspects of holography for theories with hyperscaling violation,''}
  JHEP {\bf 1206}, 041 (2012)
  \hri{1201.1905}{[hep-th]}.

\bibitem{semilocal}
S.~A.~Hartnoll and E.~Shaghoulian,
  {\em``Spectral weight in holographic scaling geometries,''}
  JHEP {\bf 1207} (2012) 078
  \hri{1203.4236}{[hep-th]}.
\\
 A.~Donos and S.~A.~Hartnoll,
  {\em``Universal linear in temperature resistivity from black hole superradiance,''}
 Phys.\ Rev.\ D {\bf 86} (2012) 124046
 \hri{1208.4102}{[hep-th]}.
\\
 R.~J.~Anantua, S.~A.~Hartnoll, V.~L.~Martin and D.~M.~Ramirez,
  {\em``The Pauli exclusion principle at strong coupling: Holographic matter and momentum space,''}
  JHEP {\bf 1303} (2013) 104
  \hri{1210.1590}{[hep-th]}.

\bibitem{Edalati:2013tma}
  M.~Edalati and J.~F.~Pedraza,
  {\em``Aspects of Current Correlators in Holographic Theories with Hyperscaling Violation,''}
Phys.\ Rev.\ D {\bf 88} (2013) 086004,
  \hri{1307.0808}{[hep-th]}.

\bibitem{RyuTaka}
 S.~Ryu and T.~Takayanagi,
 {\em``Holographic derivation of entanglement entropy from AdS/CFT,''}
  Phys.\ Rev.\ Lett.\  {\bf 96} (2006) 181602
  \hre{hep-th}{0603001};\\
 {\em``Aspects of Holographic Entanglement Entropy,''}
  JHEP {\bf 0608} (2006) 045
  \hre{hep-th}{0605073}.


\bibitem{Iqbal:2011in}
  N.~Iqbal, H.~Liu and M.~Mezei,
  {\em``Semi-local quantum liquids,''}
  JHEP {\bf 1204} (2012) 086
  \hri{1105.4621}{[hep-th]}.

\bibitem{Witten:1998zw}
  E.~Witten,
  {\em``Anti-de Sitter space, thermal phase transition, and confinement in gauge theories,''}
  Adv.\ Theor.\ Math.\ Phys.\  {\bf 2} (1998) 505
  \hre{hep-th}{9803131}.


\bibitem{Hartnoll:2011pp}
  S.~A.~Hartnoll and L.~Huijse,
  {\em``Fractionalization of holographic Fermi surfaces,''}
  Class.\ Quant.\ Grav.\  {\bf 29} (2012) 194001
 \hri{1111.2606}{[hep-th]}.

\bibitem{Adam:2012mw}
  A.~Adam, B.~Crampton, J.~Sonner and B.~Withers,
{\em``Bosonic Fractionalisation Transitions,''}
 JHEP {\bf 1301} (2013) 127
  \hri{1208.3199}{[hep-th]}.

\bibitem{Hartnoll:2012ux}
  S.~A.~Hartnoll and D.~Radicevic,
  {\em``Holographic order parameter for charge fractionalization,''}
  Phys.\ Rev.\ D {\bf 86} (2012) 066001
  \hri{1205.5291}{[hep-th]}.

\bibitem{RecBulk}
V.~E.~Hubeny,
  {\em``Extremal surfaces as bulk probes in AdS/CFT,''}
  JHEP {\bf 1207} (2012) 093
  \hri{1203.1044}{[hep-th]}.
\\
B.~Czech, J.~L.~Karczmarek, F.~Nogueira and M.~Van Raamsdonk,
  {\em``The Gravity Dual of a Density Matrix,''}
  Class.\ Quant.\ Grav.\  {\bf 29} (2012) 155009
  \hri{1204.1330}{[hep-th]}.
\\
 V.~E.~Hubeny and M.~Rangamani,
 {\em``Causal Holographic Information,''}
  JHEP {\bf 1206} (2012) 114
 \hri{1204.1698}{[hep-th]}.

\bibitem{Donos:2011bh}
  A.~Donos and J.~P.~Gauntlett,
  {\em``Holographic striped phases,''}
  JHEP {\bf 1108} (2011) 140
  \hri{1106.2004}{[hep-th]}.

\bibitem{Donos:2012yu}
  A.~Donos, J.~P.~Gauntlett, J.~Sonner and B.~Withers,
  {\em``Competing orders in M-theory: superfluids, stripes and metamagnetism,''}
  JHEP {\bf 1303} (2013) 108
  \hri{1212.0871}{[hep-th]}.

\bibitem{Donos:2013wia}
  A.~Donos,
 {\em``Striped phases from holography,''}
JHEP {\bf 1305} (2013) 059,
  \hri{1303.7211}{[hep-th]}.


\bibitem{Withers:2013loa}
  B.~Withers,
  {\em``Black branes dual to striped phases,''}
Class.\  Quant.\  Grav.\  {\bf 30} (2013) 155025,
  \hri{1304.0129}{[hep-th]}.

\bibitem{ElectronStar}
  S.~A.~Hartnoll, J.~Polchinski, E.~Silverstein and D.~Tong,
  {\em``Towards strange metallic holography,''}
  JHEP {\bf 1004} (2010) 120
  \hri{0912.1061}{[hep-th]}.
\\
 S.~A.~Hartnoll and A.~Tavanfar,
  {\em``Electron stars for holographic metallic criticality,''}
  Phys.\ Rev.\ D {\bf 83} (2011) 046003
  \hri{1008.2828}{[hep-th]}.
\\
 S.~A.~Hartnoll and P.~Petrov,
  {\em``Electron star birth: A continuous phase transition at nonzero density,''}
  Phys.\ Rev.\ Lett.\  {\bf 106} (2011) 121601
  \hri{1011.6469}{[hep-th]}.




  \bibitem{Li}
   S.~Kachru, X.~Liu and M.~Mulligan,
  {\em ``Gravity Duals of Lifshitz-like Fixed Points,''}
  Phys.\ Rev.\ D {\bf 78} (2008) 106005
  \hri{0808.1725}{[hep-th]}.


  \bibitem{taylor}
  M.~Taylor,
  {\em ``Non-relativistic holography,''}
  \hri{0812.0530}{[hep-th]}.

  \bibitem{KT}
  K.~Goldstein, S.~Kachru, S.~Prakash and S.~P.~Trivedi,
  {\em ``Holography of Charged Dilaton Black Holes,''}
  JHEP {\bf 1008} (2010) 078
  \hri{0911.3586}{[hep-th]}.



\bibitem{Gubser:2009}
  S.~S.~Gubser and A.~Nellore,
 {\em ``Ground states of holographic superconductors,''}
  Phys.\ Rev.\ D {\bf 80} (2009) 105007
  \hri{0908.1972}{[hep-th]}.


\bibitem{Horowitz:2009}
  G.~T.~Horowitz and M.~M.~Roberts,
  {\em``Zero Temperature Limit of Holographic Superconductors,''}
  JHEP {\bf 0911} (2009) 015
  \hri{0908.3677}{[hep-th]}.




  \bibitem{kasa}
  N.~Iizuka, S.~Kachru, N.~Kundu, P.~Narayan, N.~Sircar, S.~P.~Trivedi and H.~Wang,
  {\em ``Extremal Horizons with Reduced Symmetry: Hyperscaling Violation, Stripes, and a Classification for the Homogeneous Case,''}
  JHEP {\bf 1303} (2013) 126
  \hri{1212.1948}{[hep-th]}.


\bibitem{nbi}
J.~Gath, J.~Hartong, N.~A.~Obers and R.~Monteiro,
{\em``Holographic Models for Theories with Hyperscaling Violation,''}
 JHEP {\bf 1304} (2013) 159
 \hri{1212.3263}{[hep-th]}.


\bibitem{MetalIns}
 A.~Donos and S.~A.~Hartnoll,
  {\em``Interaction-driven localization in holography,''}
Nature Phys.\  {\bf 9} (2013) 649
 \hri{1212.2998}{[hep-th]}


\bibitem{Hall}
  S.~A.~Hartnoll and P.~Kovtun,
  {\em``Hall conductivity from dyonic black holes,''}
  Phys.\ Rev.\ D {\bf 76} (2007) 066001
  \hri{0704.1160}{[hep-th]}.
\\
 M.~Edalati, J.~I.~Jottar and R.~G.~Leigh,
 {\em``Transport Coefficients at Zero Temperature from Extremal Black Holes,''}
  JHEP {\bf 1001} (2010) 018
  \hri{0910.0645}{[hep-th]}.
\\
K.~Goldstein, N.~Iizuka, S.~Kachru, S.~Prakash, S.~P.~Trivedi and A.~Westphal,
  {\em``Holography of Dyonic Dilaton Black Branes,''}
  JHEP {\bf 1010} (2010) 027
  \hri{1007.2490}{[hep-th]}.


\end{thebibliography}
\end{document}